\documentclass[twocolumn,showpacs,prb,superscriptaddress,amsmath,amssymb,floatfix]{revtex4-1}
\usepackage{epsfig}
\usepackage{psfrag}
\usepackage{amsmath}
\usepackage{amsfonts}
\usepackage{amssymb}
\usepackage{graphicx}
\usepackage{bm}
\usepackage{color}

\newcommand{\bd}{\bm}

\begin{document}

\title{Functional renormalization group approach to interacting three-dimensional Weyl semimetals}
\author{Anand Sharma}
\email{sharma@itp.uni-frankfurt.de}
\affiliation{Institut f\"{u}r Theoretische Physik, Universit\"{a}t Frankfurt,  Max-von-Laue Strasse 1, 60438 Frankfurt, Germany}

\author{Arthur Scammell}
\affiliation{Institut f\"{u}r Theoretische Physik, Universit\"{a}t Frankfurt,  Max-von-Laue Strasse 1, 60438 Frankfurt, Germany}

\author{Jan Krieg}
\affiliation{Institut f\"{u}r Theoretische Physik, Universit\"{a}t Frankfurt,  Max-von-Laue Strasse 1, 60438 Frankfurt, Germany}

\author{Peter Kopietz}
\affiliation{Institut f\"{u}r Theoretische Physik, Universit\"{a}t Frankfurt,  Max-von-Laue Strasse 1, 60438 Frankfurt, Germany}
\affiliation{Department of Physics and Astronomy, University of California, Irvine, CA, 92697, USA}

\date{March 08, 2018}

\begin{abstract}

We investigate the effect of long-range Coulomb interaction on the quasiparticle properties and the dielectric function of clean three-dimensional Weyl semimetals at zero temperature using a functional renormalization group (FRG) approach. The Coulomb interaction is represented via a bosonic Hubbard-Stratonovich field which couples to the fermionic density. We derive truncated FRG flow equations for the fermionic and bosonic self-energies and for the three-legged vertices with two fermionic and one bosonic external legs. We consider two different cutoff schemes --- cutoff in fermionic or bosonic propagators --- in order to calculate the renormalized quasiparticle velocity and the dielectric function for an arbitrary number of Weyl nodes and the interaction strength. If we approximate the dielectric function by its static limit, our results for the velocity and the dielectric function are in good agreement with that of A. A. Abrikosov and S. D. Beneslavski{\u{i}} [Sov. Phys. JETP \textbf{32}, 699 (1971)] exhibiting slowly varying logarithmic momentum dependence for small momenta. We extend their result for an arbitrary number of Weyl nodes and finite frequency by evaluating the renormalized velocity in the presence of dynamic screening and calculate the wave function renormalization.

\end{abstract}

\maketitle

\section{Introduction}{\label{sec:intro}}

Three-dimensional (3D) topological semimetals are solid state materials where two bulk bands touch at either a charge neutral point, called the Dirac or Weyl node, or a line, called the nodal line, in momentum space with linear\cite{Burkov11a,Burkov16} or quadratic\cite{Kondo15} energy dispersion around it. In comparison with 3D topological insulators\cite{Hasan10}, having two-dimensional (2D) linearly dispersing surface states, 3D topological semimetals with linear energy dispersion are a distinct novel quantum state of matter since the bulk bands disperse along all three crystal momentum directions forming discrete Dirac cones in 3D momentum space. These materials are important not only due to their fundamental interest in understanding relativistic phenomena in an effective solid medium but also their macroscopic quantum properties, such as large magnetoresistance and high carrier mobility, can be promising for potential practical applications. In recent years, there have been burgeoning theoretical and experimental activities on these materials with primary focus on proposing or searching for design principles for new materials\cite{Wan11,Burkov11b,Weng14,Gibson15,Yu17}, classifying their topological nature\cite{Chiu16}, and observing exotic phenomena\cite{Hasan17,Yan17,Armitage17}. In this work, we are interested in the many-body effects due to long-range Coulomb interaction in clean 3D Weyl semimetals.

The 3D Dirac semimetals can be considered as ``dimensional extension'' of the most studied 2D topological semimetal, graphene, with fourfold degeneracy (due to crystal and time-reversal symmetry) at the Dirac node but unlike graphene this degeneracy is robust against external perturbations due to the fact that the three linearly independent momenta couple to all three Pauli matrices in the Hamiltonian. The 3D Weyl semimetals can be evolved from 3D Dirac semimetals by explicitly breaking either the spatial inversion or time-reversal symmetry. Since the electron-electron interaction plays a central role in determining the quasiparticle properties of graphene\cite{Kotov12}, it is natural to ask about the effect of many-body interaction in their 3D counterparts. This problem was first addressed and partially answered in the seminal work by Abrikosov and Beneslavski{\u{i}} for linear and quadratically dispersing 3D semimetals\cite{Abrikosov71}. In the present work, we are interested in the former case and therefore we shall only discuss linearly dispersing 3D semimetals. For any given 3D material with linear energy spectrum around a {\textit{single node}} and within {\textit{static}} effective Coulomb interaction, Abrikosov and Beneslavski{\u{i}} obtained the momentum-dependent velocity, 
\begin{equation}
\frac{v(k)}{v_F} = \left[1 + (a+b) l \right]^{\frac{a}{a+b}} ,
\label{eq:vel_AB}
\end{equation}
and the dielectric function,
\begin{equation}
\varepsilon (k) = \left[1 + (a+b) l\right]^{\frac{b}{a+b}} ,
\label{eq:de_AB}
\end{equation}
where $l = \ln\bigl(\frac{\Lambda_{0}}{k} \bigr) $ with $\Lambda_{0}$ being an ultraviolet cutoff of the order of inverse lattice spacing and $k$ is the lattice momentum. Here $a$ and $b$ are undetermined positive constants of the order of bare value of dimensionless interaction strength, 
\begin{equation}
\alpha = \frac{e^{2}}{\varepsilon_{0} v_{F}},
\label{eq:alpha}
\end{equation}
with $e$ being the bare charge, $\varepsilon_{0}$ the lattice dielectric constant, and $v_{F}$ is the Fermi velocity of the given 3D semimetal, and we use units where $\hbar = 1$. In the following, we shall not only find the constants $a$ and $b$ as a function of effective interaction strength and number of Weyl nodes but also go beyond the results of Abrikosov and Beneslavski{\u{i}} by considering any finite number of nodes and taking the frequency dependence of the dielectric function into account. Moreover, since our theory is nonperturbative in the interaction we also recover the perturbation theory results for weak interaction\cite{Hofmann15,Throckmorton15}. We employ a functional renormalization group (FRG) approach~\cite{Wetterich93,Kopietz10,Metzner12} to evaluate the renormalized quasiparticle properties (renormalized velocity, wave function renormalization) and the dielectric function for different numbers of Weyl nodes and the interaction strength of clean 3D Weyl semimetals. 

The rest of this work is organized as follows. In Sec.~\ref{sec:modes}, we introduce our low-energy effective model Hamiltonian and decouple it using a Hubbard-Stratonovich transformation. In Sec.~\ref{sec:frgapp}, we use two different cutoff strategies to derive infinite hierarchies of FRG flow equations for the irreducible vertices of our effective low-energy model. We evaluate the renormalized velocity with static and dynamically screened Coulomb interaction in Secs.~\ref{sec:ferco} and \ref{sec:bosco}, respectively, and also present the results for the static renormalized dielectric function and the wave function renormalization. In the concluding Sec.~\ref{sec:conc}, we summarize our results and suggest new directions to broaden the scope of our work.   

\section{Low-energy model}{\label{sec:modes}}

The starting point of the low-energy theory of 3D Weyl semimetals, describing a model with fermions having linear dispersion in the vicinity of charge neutral points or Weyl nodes and interacting via long-range Coulomb forces, is given by the following effective Hamiltonian,
 \begin{equation}
 {\cal{H}}_{\Lambda_{0}} =  \sum_{n} \int_{\bd{k}} \psi^{\dagger}_{n} ( \bd{k} ) ( v_{n}  \bd{\sigma} \cdot \bd{k} )
 \psi_{n} ( \bd{k} )  + \frac{1}{2}  
 \int_{\bd{q}} f_{\bd{q}} \rho_{- \bd{q}} \rho_{\bd{q}},   
 \label{eq:hamiltonian}
 \end{equation}
where $n = 1, \cdots , N_{W}$ labels the total number of Weyl nodes with nondegenerate chiralities; $v_{n} = \chi_{n} v_{F}$ is the bare velocity with $\chi_{n} = \pm$ being the chirality at the $n{\textrm{th}}$ node. It is worth mentioning that the Weyl nodes always come in pairs and the net chirality over the Brillouin zone vanishes\cite{Nielsen83}, i.e., $\sum_n \chi_n = 0$ . Thus the minimum number of Weyl nodes in a given 3D material is $N_{W} = 2$.
The operators $\psi_n ( \bd{k} )$ are two-component fermionic field operators associated with the pseudospin degrees of freedom, which in this case are the two (conduction and valence) bands touching at the $n{\textrm{th}}$ node. The 3D momentum integration is denoted by $\int_{\bd{k}} = \int \frac{ d^3 k }{ ( 2 \pi )^3}$ and is restricted to the regime $| \bd{k} | < \Lambda_0$ such that the ultraviolet cutoff $\Lambda_0$, of the order of inverse lattice spacing, is small compared to the separation between different Weyl nodes. The vector $\bd{\sigma} = ( \sigma^x, \sigma^y, \sigma^z )$, with three Pauli matrices as components, acts on the pseudospin components and the momentum $\bd{k}$ is measured relative to the nodes. The Fourier transform of the bare Coulomb interaction is given by $f_{\bd{q}} = \frac{4 \pi e^2}{\varepsilon_0 \bd{q}^2}$ and due to interest in the long-range nature of the interaction, we shall neglect any scattering processes which transfer momentum between different Weyl nodes. The Fourier component of the density operator is 
\begin{equation}
\rho_{\bd{q}} = \sum_n \int_{\bd{k}} \psi_n^{\dagger} ( \bd{k} ) \psi_n ( \bd{k} + \bd{q} ).
\end{equation}
The Euclidean action associated with the Hamiltonian (\ref{eq:hamiltonian}) is
\begin{eqnarray}
 S_{\Lambda_0} [ \bar{\psi}, \psi ]  
 & = & - \sum_{n}\int_K \psi^{\dagger}_{n} (K) [G^0_{n} ( K )]^{-1} \psi_{n}  (K)  \nonumber \\
 & + & \frac{1}{2} \int_{{Q}} f_{\bd{q}} \rho (-Q) \rho (Q),
 \label{eq:Scoupled}
 \end{eqnarray} 
where $\psi_{n} ( K )$ is a two-component Grassmann field, 
\begin{equation}
 G^0_{ n } ( K )  =  \left[  i \omega - v_{n} \bd{\sigma} \cdot \bd{k}   \right]^{-1}
 \label{eq:Gbare}
\end{equation}
being the free propagator and
\begin{equation}
\rho ( Q ) = \sum_n \int_K \psi_n^{\dagger} ( K ) \psi_n ( K + Q )
\end{equation}
is the Fourier component of the density. The multi-index label $K = ( \bd{k} , i \omega )$ represents momentum $\bd{k}$ and fermionic Matsubara frequency $i \omega$ while $Q = ( \bd{q} , i \bar{\omega} )$ is a combined label for momentum $\bd{q}$ and bosonic Matsubara frequency $i \bar{\omega}$. The integration symbols are $\int_K = \int_{\bd{k}} \int \frac{ d \omega}{2 \pi }$ and $\int_Q = \int_{\bd{q}} \int \frac{ d \bar{\omega}}{2 \pi }$ where we have considered the zero-temperature case.

In order to derive FRG flow equations, we decouple the interaction with the help of a Hubbard-Stratonovich field $\phi$ so that the regularized Euclidean action of our model becomes
\begin{eqnarray}
 S_{\Lambda_0} [\bar{\psi}, \psi , \phi ]  
 & = & - \sum_{n}\int_K \psi^{\dagger}_{n} (K) [G^0_{n} (K)]^{-1} \psi_{n}  (K)  \nonumber \\
 & + & \frac{1}{2} \int_{{Q}} \left[ f_{\bd{q}}^{-1}  \phi (-{Q}) \phi ({Q})  
 + 2 i \rho (-Q) \phi (Q) \right] ,\nonumber \\
 \label{eq:Sbare}
 \end{eqnarray} 
where the scalar Hubbard-Stratonovich field $\phi ( Q )$ couples to the density.

With the model description being set up, we follow Ref.~[\onlinecite{Kopietz10}] to derive an infinite hierarchy of FRG flow equations for the irreducible vertices of our low-energy effective model as given in Eq.~(\ref{eq:Sbare}). 

\section{FRG approach}{\label{sec:frgapp}}

Before deriving the flow equations, we have to choose a cutoff scheme by introducing a cutoff $\Lambda$ in either fermionic ($\psi$) or bosonic ($\phi$) fields by regularizing the bare fermionic or bosonic Gaussian propagators with suitable regulator functions $R_{\Lambda}^{\psi}(K)$ or $R_{\Lambda}^{\phi}(Q)$, respectively. In accordance with our previous work on graphene we shall first introduce the cutoff in fermionic propagators\cite{Bauer15} as given in Sec.~\ref{sec:ferco}, followed by cutoff in bosonic propagators\cite{Sharma16} which is explained in Sec.~\ref{sec:bosco}. The general protocol of our first approach can be summarized as follows. Once we choose the cutoff strategy, we write down the FRG flow equations for self-energies and vertices along with their diagrammatic representations. We thus arrive at an infinite hierarchy of coupled integro-differential equations. But in order to proceed further, we devise a truncation scheme based on the symmetry arguments and relevance of the vertices by simple dimensional analysis thereby retaining only marginal and relevant vertices. We finally obtain a finite small set of coupled integro-differential flow equations which we solve numerically. For the second approach, we write down the exact skeleton equation for the bosonic self-energy which can be combined with FRG flow equations for the fermionic self-energy to calculate the physical quantities of the model.     

\subsection{Fermionic cutoff scheme}{\label{sec:ferco}}

Let us introduce an infrared cutoff $\Lambda$ which prevents the propagation of electrons with momenta $| \bd{k} | < \Lambda$. With the sharp momentum cutoff, the regularized bare fermionic propagator becomes
\begin{eqnarray}
 G^0_{n, \Lambda} (K)  & = &  \left[ \left[ G^0_{ n } (K) \right]^{-1} -  R_{\Lambda}^{\psi}(K) \right]^{-1} \nonumber \\
 & = & \Theta (k  - \Lambda) G^0_{ n } (K) ,
 \label{eq:g0}
\end{eqnarray}
where the regulator function, 
\begin{eqnarray}
 R_{\Lambda}^{\psi}(K) = \left[ G^0_{ n } (K) \right]^{-1} \left[1 - \Theta^{-1} ( k  - \Lambda  ) \right],
\end{eqnarray}
is defined such that for $\Lambda \rightarrow 0$ the regulator vanishes.

Now we can write down a formally exact flow equation for the generating functional $\Gamma_{\Lambda}  [\bar{\psi}, \psi , \phi ]$ of the irreducible vertices following  Refs.~[\onlinecite{Wetterich93,Kopietz10,Metzner12}]. Note that in the limit $\Lambda \rightarrow 0$, the flowing vertices reduce to the exact irreducible vertices of the original Hamiltonian (\ref{eq:hamiltonian}). We obtain an approximate solution of the flow equations by resorting to a truncated form of $\Gamma_{\Lambda} [\bar{\psi}, \psi , \phi]$ which we choose as 
\begin{eqnarray}
\Gamma_{\Lambda} [ \bar{\psi}, \psi , \phi ] = & - & \sum_n \int_K \psi^{\dagger}_{n} (K) G^{-1}_{n, \Lambda} (K) \psi_{n} (K) \nonumber \\
& + & \frac{1}{2} \int_Q \phi (-Q) F^{-1}_{\Lambda} (Q) \phi (Q) \nonumber \\
& + & \sum_{n,b} \int_K \int_Q \Gamma^{b}_{n, \Lambda} (K+Q,K,Q) \nonumber \\
& \times & \bar{\psi}^b_n (K+Q) \psi^b_n (K)\phi (Q), \nonumber \\
\label{eq:truncate}
\end{eqnarray}
where $b = C, V$ is the conduction or valence band label, $\psi^b_n ( K )$ and $\bar{\psi}^b_n ( K )$ are the band components of spinor $\psi_n ( K )$ and its adjoint $\psi^{\dagger}_n ( K )$, and $\Gamma^{b}_{n, \Lambda} (K+Q,K,Q)$ is a three-legged vertex with two fermionic and one bosonic leg. We express the inverse scale-dependent fermionic and bosonic propagators as 
\begin{eqnarray}
 G^{-1}_{n, \Lambda} (K) & = & [G^{0}_{n, \Lambda} (K) ]^{-1}  -  \Sigma_{n, \Lambda}  ( K ),  \label{eq:ferprop} \\
 F^{-1}_{\Lambda} (Q) & = &  f_{\bd{q}}^{-1} + \Pi_{\Lambda} (Q),  \label{eq:bosprop}
\end{eqnarray}
where $ \Sigma_{n, \Lambda} ( K ),$ and $ \Pi_{\Lambda} ( Q )$ are the fermionic and bosonic self-energies, respectively, with $\Sigma_{n, \Lambda}  ( K )$ being a matrix in the band labels. Our truncation (\ref{eq:truncate}) can be justified because it contains only those vertices which are marginal and relevant as explained in the following text. 

Let us denote a vertex with $l$ external (fermionic and/or bosonic) legs as $\Gamma^{(r,s)}$ such that $l = r + s$. Then $\Gamma^{(2,1)}$ is a three-legged vertex with 2 fermionic and 1 bosonic leg. There are four vertices of this type, corresponding to the field combinations $\bar{\psi}^C \psi^C \phi$, $\bar{\psi}^V \psi^V \phi$, $\bar{\psi}^C \psi^V \phi$ and $\bar{\psi}^V \psi^C \phi$. We shall neglect the flow equations of the band-label changing three-legged vertices, with field combinations $\bar{\psi}^C \psi^V \phi$ and $\bar{\psi}^V \psi^C \phi$, since they vanish due to symmetry as it will be shown in the Appendix. The other three-legged vertex $\Gamma^{(0,3)}$ is purely bosonic with 3 bosonic legs and field combinations $\phi \phi \phi$. We can use similar notation to describe all the other higher-order vertices. Now if we keep the Gaussian part of the bare action, Eq.~(\ref{eq:Sbare}), invariant under scale transformations, we see that the scaling dimensions of the fermionic and bosonic fields are $[\psi] = -\frac{5}{2}$ and $[\phi] = -3$, respectively, as compared to $[\psi] = -2$ and $[\phi] = -2$, respectively, in case of 2D graphene. Using this fact, we find that in 3D Weyl semimetals all the four $\Gamma^{(2,1)}$ vertices and $\Gamma^{(0,4)}$ are marginal while $\Gamma^{(0,3)}$ is relevant in the RG sense. The rest of the other higher-order vertices are irrelevant at the Gaussian fixed point; i.e., they decrease under RG transformation $\Lambda \rightarrow 0$.

\begin{table}
\caption{\label{tab:table1} The relevance of the vertices in the RG sense in 3D Weyl semimetals is compared with that of 2D graphene. Though the $\Gamma^{(0,3)}$ in 3D Weyl semimetals is relevant, we will show that its flow equation vanishes due to symmetry.}
\begin{ruledtabular}
\begin{tabular}{ccc}
Vertex & 3D Weyl semimetals & 2D graphene  \\
\hline
$\Gamma^{(2,1)}$ & marginal & marginal \\
$\Gamma^{(0,3)}$ & relevant & marginal  \\
$\Gamma^{(4,0)}$ & irrelevant & irrelevant \\
$\Gamma^{(0,4)}$ & marginal & irrelevant  \\
$\Gamma^{(2,2)}$ & irrelevant & irrelevant 
\end{tabular}
\end{ruledtabular}
\end{table}

It is compelling to understand the relevance of these vertices in comparison to our previous work on 2D graphene~\cite{Bauer15} which is presented in Table~\ref{tab:table1}. Interestingly, in comparison to the 2D case, the purely bosonic three- and four-legged vertices in 3D become relevant and marginal, respectively. Such vertices play an important role in screening the charge in the 3D case but are absent in 2D graphene. Although the purely bosonic irreducible three-legged vertex is relevant, within our cutoff scheme it vanishes identically because it is absent at the initial scale, Eq.~(\ref{eq:Sbare}), and the right-hand side of its FRG flow equation vanishes by symmetry as shown in the Appendix. The four-legged vertices include processes in the particle-particle and exchange channel involving multiple particles and large momentum transfer. However, our low-energy theory is dominated by the forward scattering processes and thus we also neglect them. Moreover, since our truncation scheme is based on symmetry and relevance, it is nonperturbative in the effective dimensionless interaction strength $\alpha $ and therefore, our results are expected to hold even for large values of coupling strength.

Within our truncation and cutoff scheme, the FRG flow equations for the fermionic and bosonic self-energies are
 \begin{widetext}
 \begin{eqnarray}
 \partial_{\Lambda} \Sigma_{n,\Lambda}^{b b^{\prime}} (K) & = & \int_Q {F}_{\Lambda} (Q) \dot{G}^{b b^{\prime}}_{n,\Lambda} ( K - Q ) \Gamma^{b} _{n,\Lambda} (K,K-Q,Q) \Gamma^{b^{\prime}}_{n,\Lambda} (K-Q,K,-Q) ,
 \label{eq:flowself}
 \\ 
 \partial_{\Lambda} \Pi_{\Lambda} (Q) & = & N_s \sum_{b b^{\prime} } \sum_{n} \int_K \left[ \dot{G}_{n,\Lambda}^{b b^{\prime}} (K) G^{b^{\prime} b}_{n,\Lambda} (K-Q) +   
 {G}^{b b^{\prime}}_{n,\Lambda} (K)  \dot{G}^{b^{\prime} b }_{n,\Lambda} (K-Q)  \right]  \nonumber \\
 & & \times \Gamma^{b} _{n,\Lambda}  (K-Q,K,-Q) \Gamma^{b^{\prime}}_{n,\Lambda}  (K,K-Q,Q),
 \label{eq:flowpol}
 \end{eqnarray}
 \end{widetext}
where $N_s = 2S + 1 = 2$ is the spin degeneracy factor. Here $\Gamma^{b}_{n,\Lambda} $ represents the band-label conserving three-legged vertices, with field combinations $\bar{\psi}^C \psi^C \phi$ or $\bar{\psi}^V \psi^V \phi$, appearing in Eq.~(\ref{eq:truncate}) which satisfy the following truncated flow equation, 
 \begin{widetext}
 \begin{eqnarray} 
 \partial_{\Lambda} \Gamma^b_{n,\Lambda} (K,K-Q,Q) & = & 
 \sum_{b^{\prime}} \int_{Q^{\prime}} \Bigl\{ F_{\Lambda} ( Q^{\prime}) \Bigl[
 \dot{G}^{b b^{\prime}}_{n,\Lambda} (K-Q-Q^{\prime}) G^{ b^{\prime} b}_{n,\Lambda} (K-Q^{\prime}) +  {G}^{b b^{\prime}}_{n,\Lambda} (K-Q-Q^{\prime}) \dot{G}^{ b^{\prime} b }_{n,\Lambda} (K- Q^{\prime} ) \Bigr]
 \nonumber
 \\
 &  &  \times
  \Gamma^{b}_{n,\Lambda} (K,K-Q^{\prime},Q^{\prime}) \Gamma^{b}_{n,\Lambda} (K-Q-Q^{\prime},K-Q,-Q^{\prime}) \Gamma^{b^{\prime}}_{n,\Lambda} (K-Q^{\prime},K-Q-Q^{\prime},Q) \Bigr\} \nonumber \\
  & & - \int_{Q^{\prime}} F_{\Lambda} (Q^{\prime}) F (Q^{\prime}-Q) \dot{G}^{b b}_{n,\Lambda} (K-Q^{\prime}) \Gamma^{b}_{n,\Lambda} (K,K-Q^{\prime},Q^{\prime}) \Gamma^{b}_{n,\Lambda} (K-Q^{\prime},K-Q,Q-Q^{\prime}) \nonumber \\
  & & \times \Gamma^{(0,3)}_{n,\Lambda} (Q^{\prime},Q,Q^{\prime}-Q), 
 \label{eq:flowvertex1}
 \end{eqnarray}
 \end{widetext}
where the fermionic single-scale propagator is defined as,
 \begin{equation}
  \dot{G}_{n,\Lambda} (K) = - G_{n,\Lambda} (K) \bigl[ \partial_{\Lambda} \bigl[ G^{0}_{n,\Lambda} (K)     \bigr]^{-1} \bigr] G_{n,\Lambda} (K). 
 \end{equation}
On using the definition of the bare fermionic propagator, Eq.~(\ref{eq:g0}), and scale-dependent fermionic propagator, Eq.~(\ref{eq:ferprop}), and the Morris lemma\cite{Morris94},
 \begin{equation}
 \delta (x) f(\Theta(x)) = \delta (x) \int_{0}^{1} dt f(t),
\label{eq:morrislemma}
 \end{equation}
we get
 \begin{equation}
  \dot{G}_{n,\Lambda} (K) = -\delta (k  - \Lambda) \left[ i \omega - v_{n} \bd{\sigma} \cdot \bd{k}  - \Sigma_{n,\Lambda} (K) \right]^{-1}.
 \label{eq:gdot} 
 \end{equation}

\begin{figure}
 \centering
 \includegraphics[height=5.0cm,width=8.5cm]{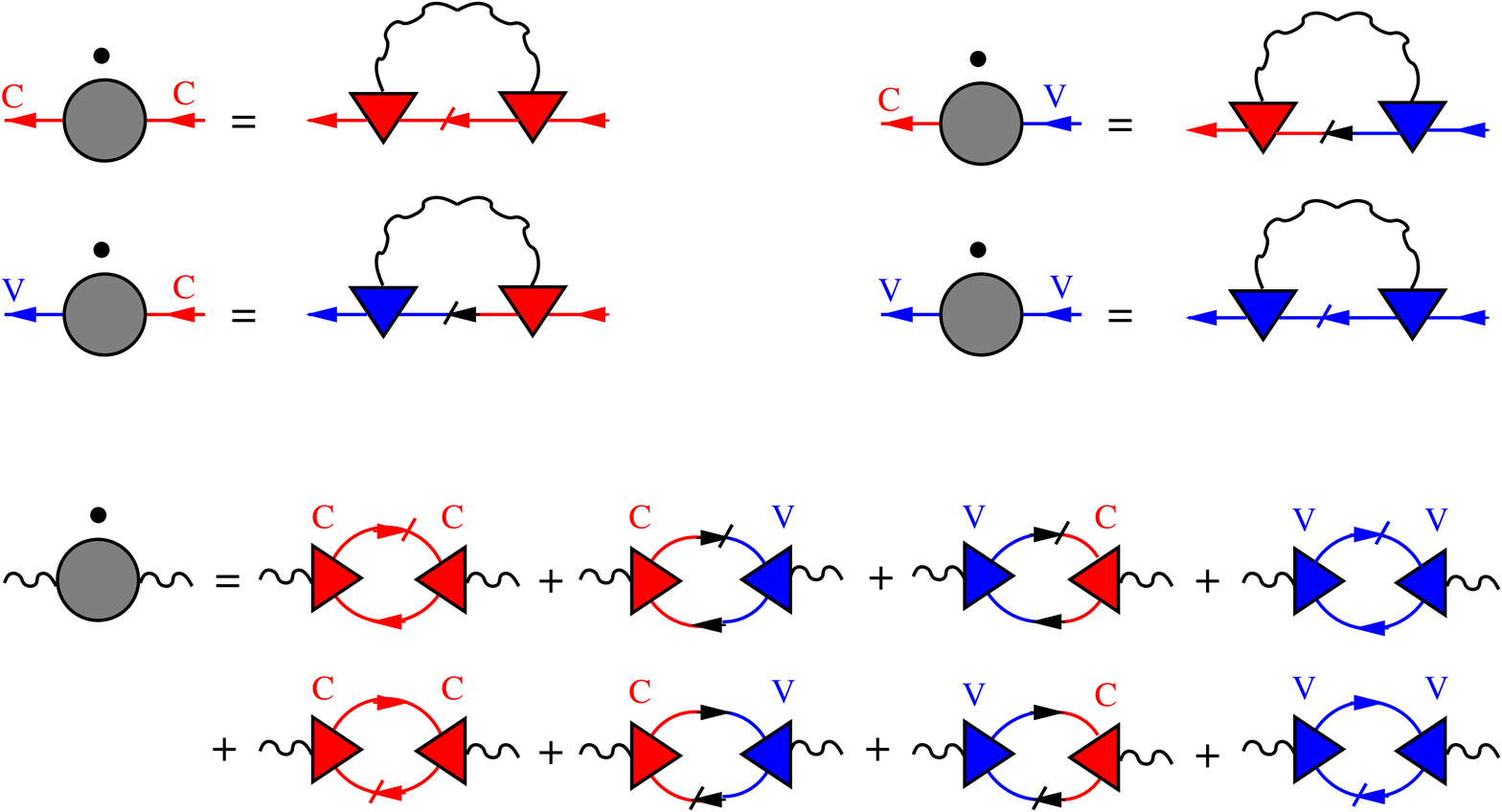}
 \caption{ Graphical representation of the FRG flow equations for the fermionic (upper panel) and bosonic (lower panel) self-energies as given in Eqs.~(\ref{eq:flowself}) and (\ref{eq:flowpol}), respectively. The dot over the symbol denotes the derivative with respect to cutoff $\Lambda$, the solid line with the arrow represents the fermionic propagator ($G^{b b^{\prime}}_{\Lambda}$), the solid line with the arrow and slash symbolizes the single-scale propagator ($\dot{G}^{b b^{\prime}}_{\Lambda}$), and the wavy line denotes the bosonic propagator ($F_{\Lambda}$). The triangles illustrate renormalized three-legged vertices with two fermionic and one bosonic leg ($\Gamma^{b}_{\Lambda}$) which carry the same band label $b$; conduction (C) shown in red and valence (V) in blue. }
 \label{fig:fcose}
 \end{figure}

\begin{figure}
 \centering
 \includegraphics[height=5.5cm,width=8.5cm]{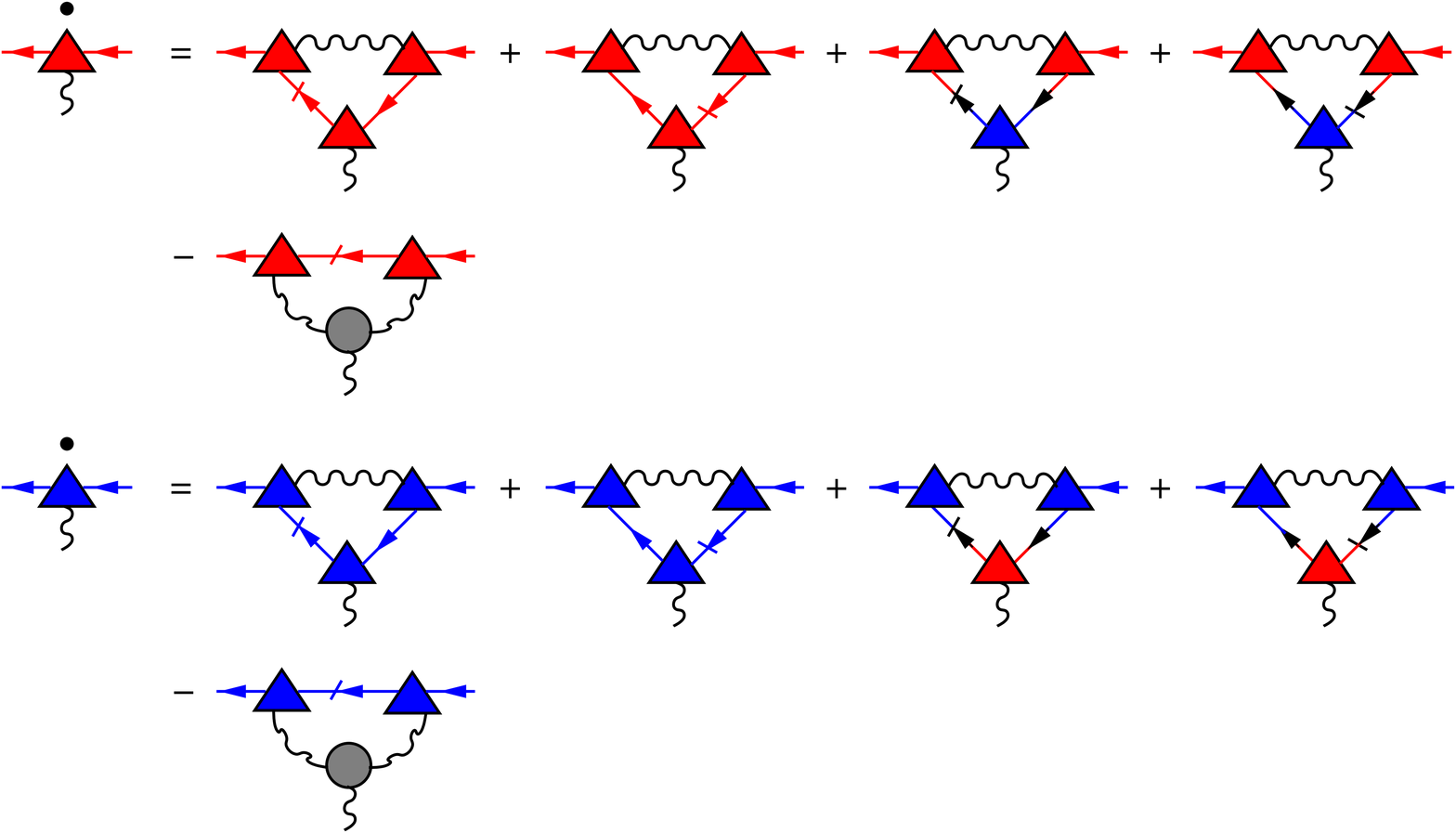}
 \caption{ Diagrammatic representation of the FRG flow equation for the three-legged vertex, Eq.~(\ref{eq:flowvertex1}), with two fermionic and one bosonic leg and carrying the same band label. The shaded circle with three wavy lines is the purely bosonic three-legged vertex. The meaning of the other symbols is the same as in Fig. \ref{fig:fcose}. }
 \label{fig:tlfv}
 \end{figure}

A graphical representation of the FRG flow equations for the fermionic and bosonic self-energies, as given in Eqs.~(\ref{eq:flowself}) and (\ref{eq:flowpol}), respectively, are shown in Fig.~\ref{fig:fcose}, while the flow equation for the same band-label three-legged vertices, Eq.~(\ref{eq:flowvertex1}), is shown in Fig.~\ref{fig:tlfv}. In the last line of Eq.~(\ref{eq:flowvertex1}), a purely bosonic three-legged vertex $\Gamma^{(0,3)}$ appears whose FRG flow equation is given by
\begin{widetext}
\begin{eqnarray}
\partial_{\Lambda} \Gamma^{(0,3)}_{n,\Lambda} (Q^{\prime},Q,Q^{\prime}-Q) &= & \sum_{b b^{\prime} b^{\prime \prime}} \int_{K} \Bigl[ 
  \dot{G}^{b b^{\prime}}_{n,\Lambda} (K) G^{ b^{\prime} b^{\prime \prime} }_{n,\Lambda} (K-Q^{\prime}+Q) G^{ b^{\prime \prime} b}_{n,\Lambda} (K-Q^{\prime}) + G^{b b^{\prime}}_{n,\Lambda} (K) \dot{G}^{ b^{\prime} b^{\prime \prime} }_{n,\Lambda} (K-Q^{\prime}+Q) G^{ b^{\prime \prime} b}_{n,\Lambda} (K-Q^{\prime})  \nonumber \\
   & + & G^{b b^{\prime}}_{n,\Lambda} (K) G^{ b^{\prime} b^{\prime \prime} }_{n,\Lambda} (K-Q^{\prime}+Q) \dot{G}^{ b^{\prime \prime} b}_{n,\Lambda} (K-Q^{\prime})  \Bigr] \Gamma^{b}_{n,\Lambda} (K-Q^{\prime},K,-Q^{\prime}) \nonumber \\
  & \times & \Gamma^{b^{\prime}}_{n,\Lambda} (K-Q^{\prime}+Q,K-Q^{\prime},Q) \Gamma^{b^{\prime \prime}}_{n,\Lambda} (K,K-Q^{\prime}+Q,Q^{\prime}-Q).
 \label{eq:flowbosvertex}  
\end{eqnarray}
\end{widetext}
In the Appendix, we shall prove that for vanishing external momentum the right-hand side of Eq.~(\ref{eq:flowbosvertex}) vanishes due to symmetry. Thus neglecting the last term in Eq.~(\ref{eq:flowvertex1}) it simplifies to
\begin{widetext}
\begin{eqnarray}
 \partial_{\Lambda} \Gamma^b_{n,\Lambda} (K,K-Q,Q) & = & 
 \sum_{b^{\prime}} \int_{Q^{\prime}} F_{\Lambda} ( Q^{\prime}) \Bigl[
 \dot{G}^{b b^{\prime}}_{n,\Lambda} (K-Q-Q^{\prime}) G^{ b^{\prime} b}_{n,\Lambda} (K-Q^{\prime}) +  {G}^{b b^{\prime}}_{n,\Lambda} (K-Q-Q^{\prime}) \dot{G}^{ b^{\prime} b }_{n,\Lambda} (K- Q^{\prime} ) \Bigr]
 \nonumber
 \\
 & & \times
  \Gamma^{b}_{n,\Lambda} (K,K-Q-Q^{\prime},Q^{\prime}) \Gamma^{b}_{n,\Lambda} (K-Q-Q^{\prime},K-Q,-Q^{\prime}) \Gamma^{b^{\prime}}_{n,\Lambda} (K-Q^{\prime},K-Q-Q^{\prime},Q).\nonumber \\ 
 \label{eq:flowvertex2}
\end{eqnarray}
\end{widetext}

In what follows we shall omit the band indices, consider a node with chirality $\chi = +1$ while the analysis is analogous for $\chi = -1$, and neglect the momentum as well as frequency dependence of the three-legged vertices thereby retaining only the marginal part, 
\begin{equation}
\Gamma_{\Lambda}^{b} (0,0,0) = i \gamma_{\Lambda} .
\label{eq:gamma}
\end{equation}
Moreover, since our interest lies in the low-energy behavior of the cutoff-dependent Green's function, we expand the self-energy to linear order in the frequency 
 \begin{equation}
 \Sigma_{\Lambda} (K) =  V_{\Lambda} (k) \bd{\sigma} \cdot \bd{k} + ( 1 - Z_{\Lambda}^{-1} ) i \omega + O( \omega^2 ),
 \label{eq:selflinear}
 \end{equation}
where $Z_{\Lambda}$ is the cutoff-dependent wave function renormalization factor. The renormalized fermionic propagator, Eq.~(\ref{eq:ferprop}), then becomes 
 \begin{equation}
 G_{\Lambda } (K)  = - \Theta (k -\Lambda) Z_{\Lambda} \frac{ i \omega +  {v}_{ \Lambda} (k) \bd{\sigma} \cdot \bd{k} }{ \omega^2 +  \xi_{\Lambda}^2 (k)},
 \label{eq:glambda}
 \end{equation}
where the renormalized velocity and renormalized energy dispersion are given by  
\begin{eqnarray}
{v}_\Lambda ( k ) & = & Z_{\Lambda} [ v_F + V_{\Lambda} ( k ) ], \label{eq:renvel} \\
\xi_{\Lambda} ( k ) & = & {v}_{\Lambda} ( k ) k . \label{eq:rendis}
\end{eqnarray}

With the above definitions of the renormalized propagators and vertices, the flow equation for the self-energy, Eq.~(\ref{eq:flowself}), becomes
 \begin{eqnarray}
 \partial_{\Lambda} \Sigma_{\Lambda} ( K ) & = & - \gamma_{\Lambda}^{2}  \int_{Q} F_{\Lambda} (Q) \dot{G}_{\Lambda} (K-Q)  \nonumber \\
 & = & - \gamma_{\Lambda}^2 Z_{\Lambda} \int \frac{d^3 q}{(2\pi)^3} \int \frac{d \bar{\omega} }{2\pi} \frac{4\pi e^2}{\varepsilon_{0} \bd{q}^2} \frac{ \delta ( |\bd{k} -\bd{q}| - \Lambda ) }{ \varepsilon_{\Lambda} (\bd{q}, i \bar{\omega})} \nonumber \\
& \times & \Biggl[ \frac{ (i \omega - i \bar{\omega}) + v_{\Lambda} (|\bd{k} -\bd{q}|) \bd{\sigma} \cdot (\bd{k} - \bd{q}) }{ (\omega - \bar{\omega})^2 + \xi_{\Lambda}^2 ( |\bd{k} -\bd{q}| ) } \Biggr] ,
 \label{eq:selfflow}
 \end{eqnarray}
where $F_{\Lambda} (Q) = \frac{f_{\bd{q}}}{\varepsilon_{\Lambda} (Q)} $ and the cutoff-dependent dielectric function is given by 
\begin{equation}
\varepsilon_{\Lambda} (Q) = 1 +  f_{\bd{q}} \Pi_{\Lambda} (Q).
\label{eq:diefun}
\end{equation}

Now in order to derive the FRG flow equations for the renormalized velocity, dielectric function, and wave function renormalization, we first set $\omega = 0$ in the FRG flow equation for the self-energy, Eq.~(\ref{eq:selfflow}), and use $\int_{-\infty}^{\infty} \frac{dx}{x^2 + a^2} = \frac{\pi}{a}$ and $\int_{-\infty}^{\infty} \frac{x dx}{x^2 + a^2} = 0$ to obtain the interaction correction to the Fermi velocity, 
\begin{align}
& \Lambda \partial_{\Lambda} V_{\Lambda} (k)\nonumber \\
& = - \gamma_{\Lambda}^{2} Z_{\Lambda} \pi \frac{\Lambda}{k} \int \frac{d^3 q}{(2\pi)^3} \frac{4\pi e^2}{\varepsilon_{0} (\bd{k} - \bd{q})^2} \frac{ \delta ( q - \Lambda ) \hat{\bd{k}} \cdot \hat{\bd{q}} }{ \varepsilon_{\Lambda} (\bd{k} - \bd{q})},
\label{eq:Vflow}  
\end{align}
where $\hat{\bd{k}}$ is the unit vector related to the quasimomentum $\bd{k}$ and we approximated $\varepsilon_{\Lambda} ( \bd{k} - \bd{q} , i \bar{\omega} ) \approx \varepsilon_{\Lambda} ( \bd{k} - \bd{q} , 0 ) \equiv \varepsilon_{\Lambda} ( \bd{k} - \bd{q} )$. The FRG flow equation of the renormalized velocity, Eq.~(\ref{eq:renvel}), then becomes 
\begin{eqnarray}
\Lambda \partial_{\Lambda} v_{\Lambda} (k) & = & \frac{v_{\Lambda} (k)}{Z_{\Lambda}} \Lambda \partial_{\Lambda} Z_{\Lambda} + \Lambda Z_{\Lambda} \partial_{\Lambda} V_{\Lambda} (k) \nonumber \\ 
& = & \eta_{\Lambda} v_{\Lambda} (k) - \gamma_{\Lambda}^2 Z_{\Lambda}^{2} \pi \frac{\Lambda}{k} \int \frac{d^3 q}{(2\pi)^3} \nonumber \\ 
& &\times  \frac{4\pi e^2}{\varepsilon_{0} (\bd{k} - \bd{q})^2} \frac{ \delta ( q - \Lambda ) \hat{\bd{k}} \cdot \hat{\bd{q}} }{ \varepsilon_{\Lambda} (\bd{k} - \bd{q})}, 
\label{eq:velflow}
\end{eqnarray}
where the flowing anomalous dimension $\eta_{\Lambda}$ is defined by
\begin{equation}
\Lambda \partial_{\Lambda} Z_{\Lambda} = \eta_{\Lambda} Z_{\Lambda}.
\label{eq:Zflow}
\end{equation}
In order to determine $ \eta_{\Lambda} $, we again consider the FRG flow equation of the self-energy, Eq.~(\ref{eq:selfflow}), but for vanishing external momentum,
\begin{eqnarray}
\eta_{\Lambda} & = & \Lambda Z_{\Lambda} \lim_{ \omega \rightarrow 0} \frac{ \partial}{\partial ( i \omega )} \partial_{\Lambda} \Sigma_{\Lambda} ( 0 , i \omega )  \nonumber \\
 & = & \gamma_{\Lambda}^2 Z_{\Lambda}^{2} \Lambda \int \frac{d^3 q}{(2\pi)^3} \int \frac{d \bar{\omega} }{2\pi} \frac{4\pi e^2}{\varepsilon_{0} \bd{q}^2} \frac{ \delta ( q - \Lambda ) }{ \varepsilon_{\Lambda} (\bd{q}, i \bar{\omega})}  \nonumber \\
 & & \times \frac{ \bar{\omega}^2 - \xi_{\Lambda}^2 (q) }{ \bigl[ \bar{\omega}^2 + \xi_{\Lambda}^2 (q) \bigr]^{2}} .
 \label{eq:eta}
\end{eqnarray}

On using Eqs.~(\ref{eq:gdot}),(\ref{eq:gamma})-(\ref{eq:glambda}) and performing the frequency integration, the FRG flow of the polarization (\ref{eq:flowpol}) becomes
 \begin{eqnarray}
\partial_{\Lambda} \Pi_{\Lambda} (Q) = & - & \gamma_{\Lambda}^2 Z_{\Lambda}^2 N_s N_W \int \frac{d^3 k}{(2\pi)^3} \Bigl[ \delta \left(k - \Lambda \right) \nonumber \\ 
& \times & \Theta \left(|\bd{k} - \bd{q}|  - \Lambda \right)  
 + \Theta  \left(k - \Lambda \right) \delta \left(|\bd{k} - \bd{q}|  - \Lambda \right) \Bigr] \nonumber \\ 
 & \times & \biggl[ 1 - \frac{ \bd{k} \cdot (\bd{k} - \bd{q}) }{ k|\bd{k} - \bd{q}| } \biggr] \nonumber \\ 
 & \times & \frac{ \xi_{\Lambda} (k ) +  \xi_{\Lambda} (|\bd{k} - \bd{q}|) }{ [ \xi_{\Lambda} (k) + \xi_{\Lambda} (|\bd{k} -\bd{q}|) ]^2 + \bar{\omega}^2} ,
 \label{eq:polflow}
 \end{eqnarray}
which yields the flowing dielectric function, 
\begin{equation}
\Lambda \partial_{\Lambda} \varepsilon_{\Lambda} (Q) = \Lambda f_{\bd{q}} \partial_{\Lambda} \Pi_{\Lambda} (Q).
\label{eq:dieflow}
\end{equation}

Finally, we consider the FRG flow equation of the momentum- and frequency-independent part of the same band-label three-legged vertex given by
\begin{equation}
\partial_{\Lambda} \gamma_{\Lambda} = -\gamma_{\Lambda}^3 \int_{Q} F_{\Lambda} (Q) \textrm{Tr} [ \dot{G}_{\Lambda} (Q) G_{\Lambda}(Q) ],
\label{eq:vertflow}
\end{equation}
where Tr is the trace over the product of fermionic propagators. On writing the fermionic single-scale propagator as
\begin{equation}
\dot{G}_{\Lambda} (Q) = -\frac{ \delta (q - \Lambda) \bigl[ G^{0} (Q) \bigr]^{-1} }{ \bigl[ \bigl[ G^{0}(Q) \bigr]^{-1} - \Theta (q - \Lambda) \Sigma_{\Lambda} (Q) \bigr]^{2}} ,
\label{eq:singlescaleprop}
\end{equation}
where $G^{0}(Q)$ is the bare propagator, Eq.~(\ref{eq:Gbare}), for $\chi = 1$ and using Eq.~(\ref{eq:glambda}) along with the Morris lemma, Eq.~(\ref{eq:morrislemma}), we get
\begin{equation}
\partial_{\Lambda} \gamma_{\Lambda} = \gamma_{\Lambda}^3 \int_{Q} F_{\Lambda} (Q) \frac{\delta (q - \Lambda)}{2} \textrm{Tr} \bigl[ [ G^{0} (Q) ]^{-1} - \Sigma_{\Lambda} (Q) \bigr]^{-2}.
\end{equation}
We now utilize the low-energy expansion for the self-energy, Eq.~(\ref{eq:selflinear}), and perform the trace to obtain
\begin{equation}
\partial_{\Lambda} \gamma_{\Lambda} = -\gamma_{\Lambda}^3 Z_{\Lambda}^{2} \int_{Q} F_{\Lambda} (Q) \delta (q - \Lambda) \frac{ \bar{\omega}^2 - \xi_{\Lambda}^2 (q) }{ \bigl[ \bar{\omega}^2 + \xi_{\Lambda}^2 (q) \bigr]^{2}},
\end{equation}
which can be written as 
\begin{equation}
\Lambda \partial_{\Lambda} \gamma_{\Lambda} = - \eta_{\Lambda} \gamma_{\Lambda},
\label{eq:vertflow2}
\end{equation}
where the flowing anomalous dimension $\eta_\Lambda$ is given in Eq.~(\ref{eq:eta}). This implies $\partial_{\Lambda} (\gamma_{\Lambda} Z_{\Lambda})  = 0$ and along with the initial condition $ \gamma_{\Lambda_{0}} = Z_{\Lambda_{0}} = 1 $ we obtain $\gamma_{\Lambda}  Z_{\Lambda} =1$ which satisfies the Ward identity related to the charge conservation. Since the combination $ \gamma_{\Lambda} Z_{\Lambda} $ appears in the flow equations for the velocity (\ref{eq:velflow}), the anomalous dimension (\ref{eq:eta}), and the polarization (\ref{eq:polflow}), there is an exact cancellation between wave function renormalization and vertex corrections in these flow equations such that neither $\gamma_{\Lambda}$ nor $Z_{\Lambda}$ appears separately on the right hand of the flow equations. Therefore, instead of four FRG flow equations we are left with the flow equations for renormalized velocity $v_{\Lambda}(k)$, wave function renormalization $Z_{\Lambda}$, and the dielectric function $\varepsilon_{\Lambda} (Q)$ which forms a closed set of three coupled integro-differential equations that can be solved numerically to determine the momentum- and frequency-dependent quasiparticle properties of the model system with the number of Weyl nodes $N_W$ and the strength of the interaction $\alpha$, Eq.~(\ref{eq:alpha}), being the model parameters.

Now before we solve the closed system of dynamical FRG flow equations for $v_\Lambda ( k )$, $Z_{\Lambda}$ and $\varepsilon_{\Lambda} ( \bd{q} , i \bar{\omega} )$, we examine the static (frequency-independent) FRG flow equations. This further reduces the number of coupled integro-differential equations, from three to two, since in this approximation $Z_{\Lambda} = \gamma_{\Lambda} = 1$. Note that, in this approximation, the renormalization of the Fermi velocity is still nonperturbatively taken into account. The flow equations, (\ref{eq:velflow}) and (\ref{eq:polflow}), get further simplified on utilizing the Ward identity $\gamma_{\Lambda}  Z_{\Lambda} =1$ and employing prolate spheroidal coordinates in Eq.~(\ref{eq:polflow}),
\begin{eqnarray}{\label{eq:sphprocor}}
k_{x} = \frac{|{\bf{q}}|}{2} {\rm{sinh}}\mu \hspace{0.1cm} {\rm{sin}}\varphi \hspace{0.1cm} {\rm{cos}}\nu ,\nonumber \\
k_{y} = \frac{|{\bf{q}}|}{2} {\rm{sinh}}\mu \hspace{0.1cm} {\rm{sin}}\varphi \hspace{0.1cm} {\rm{sin}}\nu ,\\
k_{z} = \frac{|{\bf{q}}|}{2} ({\rm{cosh}}\mu \hspace{0.1cm} {\rm{cos}}\varphi - 1), \nonumber 
\end{eqnarray}
\\
with $\mu \in [0,\infty)$, $\varphi \in [0,\pi]$, $\nu \in [0,2\pi)$ and
\begin{equation}
d^{3}k = \frac{|{\bf{q}}|^{3}}{8} \hspace{0.1cm} {\rm{sinh}}\mu \hspace{0.1cm} {\rm{sin}}\varphi \hspace{0.1cm} ({\rm{cosh}}^{2}\mu - {\rm{cos}}^{2}\varphi ) \hspace{0.1cm} d\mu \hspace{0.1cm} d\varphi \hspace{0.1cm} d\nu.
\end{equation}
Thus, for Eqs.~(\ref{eq:velflow}) and (\ref{eq:dieflow}) we finally obtain
 \begin{widetext}
 \begin{eqnarray}
 \Lambda \partial_{\Lambda} \biggl[\frac{v_{\Lambda} ( k )}{v_{F}}\biggr] & = &
  - \frac{\alpha}{2} \frac{\Lambda}{k} \int_0^{\pi} \frac{d \varphi}{\pi} \frac{ \sin \varphi \cos \varphi}{ \Bigl[ 1 - 2 \bigl(\frac{k}{\Lambda}\bigr) \cos \varphi + \bigl(\frac{k}{\Lambda}\bigr)^2 \Bigr] }
 \frac{1}{\varepsilon_{\Lambda} \bigl( \sqrt{ \Lambda^2 - 2 k \Lambda \cos \varphi + k^2 } \bigr)},
 \label{eq:flowvel} 
 \\
  \Lambda \partial_{\Lambda} \varepsilon_{\Lambda} ( q ) & = &
  -  \alpha N_s N_W  
 \int_0^{\pi/2} \frac{d \varphi}{\pi} 
 \frac{ \sin^3 \varphi  \hspace{0.3cm} \Theta ( 1 + \frac{q}{2 \Lambda} \cos \varphi 
  - \frac{q}{2 \Lambda}) } { \Bigl\{ \Bigl[\frac{v_{\Lambda} ( \Lambda )}{v_{F}}\Bigr] + 
 \Bigl( 1 + \bigl( \frac{q}{\Lambda} \bigr) \cos \varphi \Bigr) \Bigl[ \frac{v_{\Lambda} ( \Lambda + q  \cos \varphi )}{v_{F}} \Bigr] \Bigr\} }.
 \label{eq:flowdie}
 \end{eqnarray}
 \end{widetext} 

 \begin{figure}
 \centering
 \includegraphics[height=7.0cm,width=8.5cm]{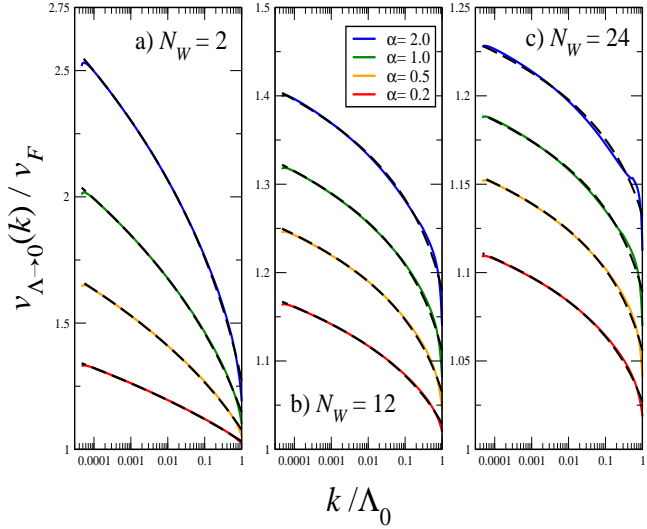} 
 \caption{ Static renormalized velocity in the limit of vanishing cutoff as a function of quasimomentum $k$ for (a) $N_W$ = 2, (b) $N_W$ = 12, and (c) $N_W$ = 24 and $\alpha = 0.2, 0.5, 1.0, 2.0 $. The broken black line is a fit given in the form of Eq. (\ref{eq:fitvel}). }
 \label{fig:flowvel}
 \end{figure}
We solve the coupled integro-differential equations, (\ref{eq:flowvel}) and (\ref{eq:flowdie}), for the momentum- and cutoff-dependent renormalized velocity $v_{\Lambda} ( k )$ and static dielectric function $\varepsilon_{\Lambda} ( q )$ for different values of $N_W$ and $\alpha$. There have been a substantial number of theoretical proposals and a few experimental observations on 3D Weyl semimetals\cite{Xu11,Halasz12,Singh12,Bulmash14,Liu14,Wang16,Liu17,Lv15a,Xu15a,Lu15,Lv15b,Xu15b,Xu15c} but due to lack of complete information on the values of $N_W$ and $\alpha$, we shall consider $N_W$ and $\alpha$ as parameters and perform the calculations for $N_W = 2, 12, 24$ and $\alpha = 0.2, 0.5, 1.0, 2.0$. As already mentioned, our theory is valid for any finite value of interaction strength $\alpha$ and is not restricted to the perturbative weak-coupling regime. In fact we shall show that we can recover the static nonperturbative result of Abrikosov and Beneslavski{\u{i}} from which one can obtain the perturbative solution of Throckmorton {\textit{et al}}\cite{Throckmorton15}. Moreover, we shall also demonstrate that the constants $a$ and $b$, appearing in the theory of Abrikosov and Beneslavski{\u{i}}\cite{Abrikosov71} as given in Eqs.~(\ref{eq:vel_AB}) and (\ref{eq:de_AB}), can be derived in a straightforward manner from Eqs.~(\ref{eq:flowvel}) and (\ref{eq:flowdie}). 

 \begin{figure}
 \centering
 \includegraphics[height=7.0cm,width=8.5cm]{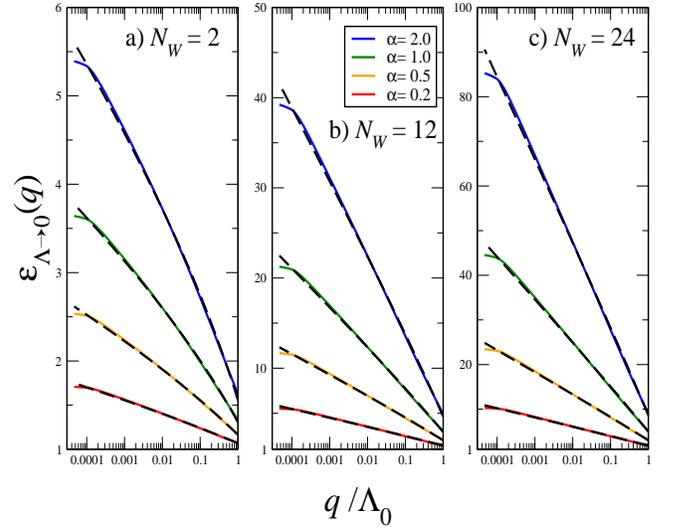}
 \caption{ Static renormalized dielectric function in the limit of vanishing cutoff as a function of quasimomentum $q$ for a) $N_W$ = 2, b) $N_W$ = 12 and c) $N_W$ = 24 and $\alpha = 0.2, 0.5, 1.0, 2.0 $. The broken black line is a fit given in the form of Eq. (\ref{eq:fitdie}).}
 \label{fig:flowdie} 
 \end{figure}
In Figs. \ref{fig:flowvel} and \ref{fig:flowdie} we show the results for the physical renormalized velocity and the static dielectric function, respectively, which is defined in the limit $\Lambda \rightarrow 0$.  We find that both the renormalized velocity and the dielectric function diverge logarithmically for vanishing momentum. Though the logarithmic divergence of the velocity is similar to the case of 2D graphene~\cite{Bauer15,Sharma16}, the bare long-range Coulomb interaction gets strongly screened in the case of 3D Weyl semimetals because the polarization bubble in 3D diverges logarithmically and so does the dielectric function, Eq.~(\ref{eq:diefun}). This is in contrast with the case of 2D graphene where in the static limit the dielectric function becomes unity with vanishing momentum since the polarization bubble in 2D is proportional to the momentum. The general qualitative behavior for the 3D Weyl semimetals is such that for a fixed number of Weyl nodes $N_W$, the renormalized velocity and the dielectric function increases with increasing interaction $\alpha$ while for a given fixed coupling strength $\alpha$, the renormalized velocity decreases and the dielectric function increases with increasing $N_W$. 

In order to have a quantitative understanding of the low-momentum behavior of the renormalized velocity and dielectric function, let us proceed by simplifying the coupled integro-differential FRG flow  equations (\ref{eq:flowvel}) and (\ref{eq:flowdie}). If we ignore the momentum dependence, by taking the limit of vanishing momentum $ k \rightarrow 0$ in  Eq.~(\ref{eq:flowvel}), we obtain 
 \begin{eqnarray}
 \Lambda \partial_{\Lambda} \biggl[ \frac{ v_{\Lambda}}{ v_{F}} \biggr] & = & - \frac{\alpha}{2} \frac{\Lambda}{k} \frac{1}{\varepsilon_{\Lambda}} \int_0^{\pi} \frac{d \varphi}{\pi} \frac{ \sin \varphi \cos \varphi}{ \Bigl[ 1 - 2 \bigl( \frac{k}{\Lambda} \bigr) \cos \varphi + \bigl( \frac{k}{\Lambda} \bigr)^2 \Bigr] } \nonumber \\
 & = & - \frac{2 \alpha}{3 \pi} \frac{1}{\varepsilon_{\Lambda}},
 \label{eq:flowvel2} 
 \end{eqnarray}
where the logarithm appearing in the trivial angular integral has been expanded such that ln$(1+x) \approx x - \frac{x^2}{2} + \frac{x^3}{3}$ for small $x = \frac{k}{\Lambda}$. On introducing the RG flow parameter $l =$ ln$\bigl(\frac{\Lambda_{0}}{\Lambda} \bigr)$ and with $ \Lambda \partial_{\Lambda} = -\partial_{l} $ we obtain the RG flow equation for the renormalized velocity as
\begin{equation}
\partial_{l} \biggl[ \frac{v_{l}}{v_F} \biggr] = \frac{a}{\varepsilon_{l}}
 \label{eq:flowvel3} 
\end{equation}
with 
\begin{equation}
a = \frac{2 \alpha}{3 \pi}
\end{equation}
being a dimensionless constant and we have renamed $\bigl[ \frac{v_{\Lambda}}{v_{F}} \bigr] \rightarrow \bigl[ \frac{v_{l}}{v_{F}} \bigr]$ and $\varepsilon_{\Lambda} \rightarrow \varepsilon_{l}$. Similarly if we take the limit of vanishing momentum $q \rightarrow 0$ in Eq.~(\ref{eq:flowdie}) we get
 \begin{eqnarray}
 & &  \Lambda \partial_{\Lambda} \varepsilon_{\Lambda} \nonumber \\
 & = & - \frac{\alpha N_s  N_W}{\bigl[\frac{v_{\Lambda}}{v_{F}} \bigr] }  \int_0^{\pi/2} \frac{d \varphi}{2 \pi} 
 \frac{ \sin^3 \varphi  \hspace{0.3cm} \Theta ( 1 + \frac{q}{2 \Lambda} \cos \varphi 
  - \frac{q}{2 \Lambda}) } { \bigl( 1 + \bigl(\frac{q}{2\Lambda} \bigr) \cos \varphi \bigr) } \nonumber \\
  & = & - \frac{\alpha N_s N_W}{3 \pi} \frac{1}{\bigl[\frac{v_{\Lambda}}{v_{F}} \bigr]},
  \label{eq:flowdie2} 
 \end{eqnarray}
which yields the RG flow equation for the renormalized dielectric function
\begin{equation}
\partial_{l} \varepsilon_{l} = \frac{b}{ \bigl[\frac{v_{l}}{v_{F}} \bigr]},
\label{eq:flowdie3} 
\end{equation}
where we have defined the dimensionless constant
\begin{equation}
b = \frac{\alpha N_s N_W}{3 \pi} = \frac{a N_s N_W}{2}.
\end{equation}
From Eqs.~(\ref{eq:flowvel3}) and (\ref{eq:flowdie3}) we see that
\begin{equation}
\partial_{l} \biggl\{ \biggl[\frac{v_{l}}{v_{F}} \biggr] \varepsilon_{l} \biggr\} = a + b,
\end{equation}
so that
\begin{equation}
\varepsilon_{l} = \frac{1 + (a + b)l}{\bigl[\frac{v_{l}}{v_{F}} \bigr]}.
\label{eq:flowdie4} 
\end{equation}
On substituting Eq.~(\ref{eq:flowdie4}) into Eq.~(\ref{eq:flowvel3}) and on integrating we obtain
\begin{eqnarray}
\biggl[\frac{v_{l}}{v_{F}} \biggr] & = & [1 + (a + b)l]^{\frac{a}{a+b}} \nonumber \\
& = & \Biggl[1 + \frac{\alpha\left(2+N_s N_W\right)}{3 \pi}l \Biggr]^{\frac{2}{2+N_s N_W}},
\label{eq:flowvel4} 
\end{eqnarray}
while substituting Eq.~(\ref{eq:flowvel4}) into Eq.~(\ref{eq:flowdie4}) we get
\begin{eqnarray}
\varepsilon_{l} & = & [1 + (a + b)l]^{\frac{b}{a+b}} \nonumber \\
& = & \Biggl[1 + \frac{\alpha(2+N_s N_W )}{3 \pi}l \Biggr]^{\frac{N_s N_W}{2+N_s N_W}}.
\label{eq:flowdie5} 
\end{eqnarray}
Thus we derive the renormalized velocity (\ref{eq:flowvel4}) and the static dielectric function (\ref{eq:flowdie5}) which is one of the main results of this work. It provides a more rigorous quantitative understanding of the behavior of the quasiparticle properties in 3D Weyl semimetals. It can be shown using the FRG method\cite{Bauer15} that $v_{\Lambda \rightarrow 0}(k) $ can be obtained from $ v_{\Lambda}(k=0)$ by setting $\Lambda \approx k$; hence we can make the substitution $l=$ ln$\bigl(\frac{\Lambda_0}{k}\bigr)$, so that Eqs. (\ref{eq:flowvel4}) and (\ref{eq:flowdie5}) can be expressed in terms of $k$ as in Eqs. (\ref{eq:vel_AB}) and (\ref{eq:de_AB}). In comparison with the theory of Abrikosov and Beneslavski{\u{i}}, we not only obtain the values for the undetermined constants $a = \frac{2\alpha}{3 \pi}$ and $b = \frac{\alpha N_s N_W}{3 \pi}$, as mentioned in Eqs.~(\ref{eq:vel_AB}) and (\ref{eq:de_AB}), but also recover their theory for the case of the {\textit{single node}}, $N_W = 1$ and $N_s = 1$. However, we note that since the Weyl nodes always come in pairs\cite{Nielsen83}, $N_W$ should always be an even number as defined in Sec.~\ref{sec:modes}. Moreover, since our result is valid for any finite effective interaction strength $\alpha$, we recover the perturbative results of Throckmorton {\textit{et al.}}\cite{Throckmorton15} on expanding Eqs.~(\ref{eq:flowvel4}) and (\ref{eq:flowdie5}) for small coupling $\alpha$.

\begin{figure}
 \centering
 \includegraphics[height=6.5cm,width=8.5cm]{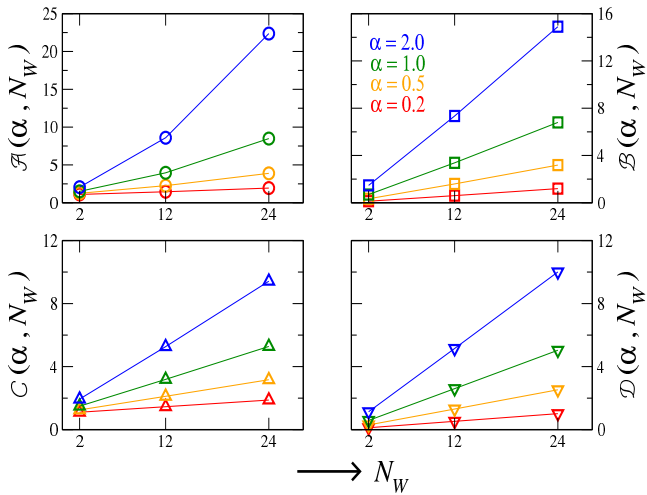}
 \caption{ The values for $\mathcal{A}(\alpha,N_W)$ and $\mathcal{B}(\alpha,N_W)$, from Eq.~(\ref{eq:fitvel}), are shown in the upper panel as {\scriptsize{}$\bigcirc$} and $\square$, respectively, and obtained as the fitting parameters to the numerical solution of Eq.~(\ref{eq:flowvel}) which is shown in Fig.~\ref{fig:flowvel}. In the lower panel, the values for $\mathcal{C}(\alpha,N_W) $ and $\mathcal{D}(\alpha,N_W) $, from Eq.~(\ref{eq:fitdie}), are shown as {\large{}$\vartriangle$} and {\large{}$\triangledown$}, respectively, and obtained as the fitting parameters to the numerical solution of Eq.~(\ref{eq:flowdie}) which is shown in Fig.~\ref{fig:flowdie}. }
 \label{fig:abcd} 
 \end{figure}

On assuming that the power in Eqs.~(\ref{eq:flowvel4}) and (\ref{eq:flowdie5}) gives the correct quantitative behavior, we fit the numerically obtained results for the renormalized velocity and the  dielectric function using the following formula,
\begin{eqnarray}
v_{l} & = & \Bigl[\mathcal{A}(\alpha,N_W) + \mathcal{B}(\alpha,N_W) l \Bigr]^{\frac{1}{1+N_W}}, \label{eq:fitvel} \\
\varepsilon_{l} & = & \Bigl[\mathcal{C}(\alpha,N_W) + \mathcal{D}(\alpha,N_W) l \Bigr]^{\frac{N_W}{1+N_W}}, \label{eq:fitdie} 
\end{eqnarray}
where we have considered the spin degeneracy factor $N_s = 2$ and the dimensionless variables $\mathcal{A}, \mathcal{B}, \mathcal{C} $ and $\mathcal{D}$ are functions of $\alpha$ and $N_W$. The values obtained from the fit for $\mathcal{A} $ and $\mathcal{B}$ are shown in the upper panel while those for $\mathcal{C} $ and $\mathcal{D}$ are exhibited in the lower panel of Fig.~\ref{fig:abcd}. We find that for a fixed number of Weyl nodes $N_W$ (strength of effective interaction $\alpha$), the values of $\mathcal{A}, \mathcal{B}, \mathcal{C} $, and $\mathcal{D}$ increase with increasing $\alpha$ ($N_W$). Note that this is compatible with the fact that for a given $\alpha$ and $N_W$, the values for $\mathcal{B}$ and $\mathcal{D}$ can be compared with $a + b = \frac{2\alpha(1+N_W)}{3 \pi} $ for $N_s = 2$ and therefore it is proportional to $\alpha$ and $N_W$. Since, within a given accuracy, the fitting is more accurate for the renormalized dielectric function $\varepsilon_l$, the values of $\mathcal{D}$ are closer to $a + b = \frac{2\alpha(1+N_W)}{3 \pi} $ than those of $\mathcal{B}$.

As seen from Fig.~\ref{fig:flowvel}, the fit to the numerical solution of the renormalized velocity, Eq.~(\ref{eq:flowvel}), agrees very well for weak coupling and small $N_W$ but starts to deviate for large values of $\alpha$ and $N_W$. The deviation is not that significant for the renormalized dielectric function, Eq.~(\ref{eq:flowdie}), for large values of $N_W$ and $\alpha$ as is obvious from Fig.~\ref{fig:flowdie}. It is clear that the non-trivial task of finding the numerical solution of the static coupled FRG flow equations for the renormalized velocity and the dielectric function is affected by the choice of the numerical method of integration, interpolation, and grid size used in the calculation. In particular, for large values of $\alpha$ and $N_W$ the dielectric function increases rapidly which raises numerical difficulties for the interpolation function of a given grid size. Such a scenario becomes even more demanding in the case of dynamical FRG flow equations for $v_\Lambda ( k )$, $\varepsilon_{\Lambda} ( \bd{q} , i \bar{\omega} )$, and $Z_{\Lambda}$ because of the additional frequency integral. Therefore, instead of solving the full dynamical solution for the renormalized velocity, dielectric function and the wavefunction renormalization we shall resort to another approach which is explained in the following Sec.~\ref{sec:bosco}.
 
\subsection{Bosonic cutoff scheme}{\label{sec:bosco}}

The details of this scheme are presented in our earlier paper on 2D graphene~\cite{Sharma16}. Therefore for the sake of brevity, here we just mention important points concerning the 3D Weyl semimetals and wherever necessary we shall compare it with the 2D case. One of the advantages of this bosonic cutoff scheme is that it can be combined with Dyson-Schwinger equations in the bosonic sector to obtain a closed FRG flow equation for the fermionic self-energy and therefore providing direct access to the required quasiparticle properties of the system. 

In this approach, we introduce a cutoff only in the bosonic propagators. For our purpose it is sufficient to work with a sharp momentum cutoff which restricts the momentum transferred by the bosonic field to the regime $q > \Lambda$. Thus the regularized free bosonic propagator becomes
\begin{eqnarray}
 F^0_{\Lambda} (Q)  & = &  \left[  f_{\bd{q}}^{-1}  -  R_{\Lambda}^{\phi}(Q) \right]^{-1} \nonumber \\
 & = & \Theta (q  - \Lambda) f_{\bd{q}} ,
\end{eqnarray}
where the regulator function, 
\begin{eqnarray}
 R_{\Lambda}^{\phi}(Q) = f_{\bd{q}} ^{-1} \left[1 - \Theta^{-1} ( q  - \Lambda  ) \right],
 \vspace{0.5cm}
\end{eqnarray}
vanishes for $\Lambda \rightarrow 0$. Now the exact flow equation for the fermionic self-energy in this scheme can be derived from the general hierarchy of FRG flow equations, Ref.~[\onlinecite{Kopietz10}], and is given by
 \begin{widetext}
 \begin{eqnarray}
 \partial_{\Lambda} \Sigma_{n,\Lambda}^{b b^{\prime}} (K) = \sum_{b_{1} b_{2}} \int_Q \dot{F}_{\Lambda} (Q) G^{b_{1} b_{2}}_{n,\Lambda} ( K - Q ) \Gamma^{b b_{1} \phi}_{n,\Lambda} (K,K-Q,Q) \Gamma^{b_{2} b^{\prime} \phi}_{n,\Lambda} (K-Q,K,-Q) + \frac{1}{2} \int_Q \dot{F}_{\Lambda} (Q) \Gamma^{(2,2)}_{n,\Lambda} (K,K,Q,-Q), \nonumber \\
 \label{eq:flowself2}
 \end{eqnarray}
 \end{widetext}
where the fermionic propagator is related to the fermionic self-energy by the Dyson equation (\ref{eq:ferprop}). The external legs connected to the three-legged vertex $\Gamma^{b b^{\prime} \phi} (K,K-Q,Q)$ correspond to the fields $\bar{\psi}^b (K), \psi^{b^{\prime}} (K-Q)$, and $\phi (Q)$. There are four types of such vertices as explained in Sec.~\ref{sec:ferco} but in the following we shall only consider the band-label conserving three-legged vertices since the RG flow equation of the band-label changing vertices vanishes due to symmetry. The four-legged vertex $ \Gamma^{(2,2)} (K,K,Q,-Q) $ has two fermionic legs $\bar{\psi}^b (K)$ and $\psi^{b^{\prime}} (K)$ and two bosonic legs $\phi (Q)$ and $\phi (-Q)$. The bosonic single-scale propagator is given by
 \begin{eqnarray}
  \dot{F}_{\Lambda} (Q) & = & - F_{\Lambda}^{2} (Q)  \partial_{\Lambda} \bigl[ F^0_{\Lambda} (Q) \bigr]^{-1}    \nonumber \\
  & =  &  -\delta (q - \Lambda) \left[ f_{\bd{q}} ^{-1}  + \Pi_{\Lambda} (Q) \right]^{-1},
  \label{eq:fdot}
 \end{eqnarray}
where $\Pi_{\Lambda} (Q) $ is the irreducible particle-hole bubble or the bosonic self-energy which we shall write using the exact Dyson-Schwinger equation,
 \begin{eqnarray} 
 \Pi_{\Lambda} (Q) & = & i N_s \sum_{b b^{\prime} } \sum_{n} \int_K {G}_{n,\Lambda}^{b b^{\prime}} (K) G^{b^{\prime} b}_{n, \Lambda} (K-Q) \nonumber \\
 & & \times \Gamma^{b^{\prime} b^{\prime} \phi}_{n,\Lambda} (K,K-Q,Q).
 \label{eq:skeleton}
 \end{eqnarray}
Note that following Refs.~[\onlinecite{Sharma16,Schuetz05}], we have closed the RG flow equation by means of the skeleton equation for the polarization bubble instead of deriving the FRG flow equation for it. 

Now we can obtain a closed set of equations with the knowledge of the FRG flow equations for the three- and four-legged vertices along with Eq.~(\ref{eq:flowself2}). But it is well known that the flow equations of these three- and four-legged irreducible vertices are of hierarchical nature generating higher-order irreducible vertices for which one has to again write down their flow equations. In order to keep the approach tractable with less numerical effort and still obtain sensible results, we shall use the truncation strategy as used in the previous Sec.~\ref{sec:ferco} which was based on the classification of vertices according to symmetry and relevance. We shall retain only those vertices which are marginal or relevant and are finite at the initial RG scale implying that we can neglect the mixed four-legged vertex since it is irrelevant (see Table~\ref{tab:table1}). Moreover, the band-label changing three-legged vertices are not only absent at the initial scale but, as mentioned earlier, their flow equation vanishes due to symmetry (see Appendix) and therefore we shall also neglect them. Thus, we finally arrive at the following flow equation for the fermionic self-energy,
 \begin{equation}
 \partial_{\Lambda} \Sigma_{\Lambda} ( K ) = - \gamma_{\Lambda}^2 \int_Q  \dot{F}_{\Lambda} (Q) G_{\Lambda} (K-Q),
 \label{eq:selftrunc}
 \end{equation}
where we have omitted the band labels, considered a node with chirality $\chi = +1$, and have retained only the marginal part of the three-legged vertex, see Eq.~(\ref{eq:gamma}), by neglecting its momentum and frequency dependence. In this approximation the exact skeleton equation (\ref{eq:skeleton}) becomes 
 \begin{equation}
 \Pi_{\Lambda} (Q) =  - \gamma_{\Lambda} N_s N_W \int_K  \textrm{Tr} [ G_{\Lambda} (K) G_{\Lambda}(K-Q) ].
 \label{eq:skeleton2}
 \end{equation}
 \begin{figure}
 \centering
 \includegraphics[height=4.0cm,width=8.0cm]{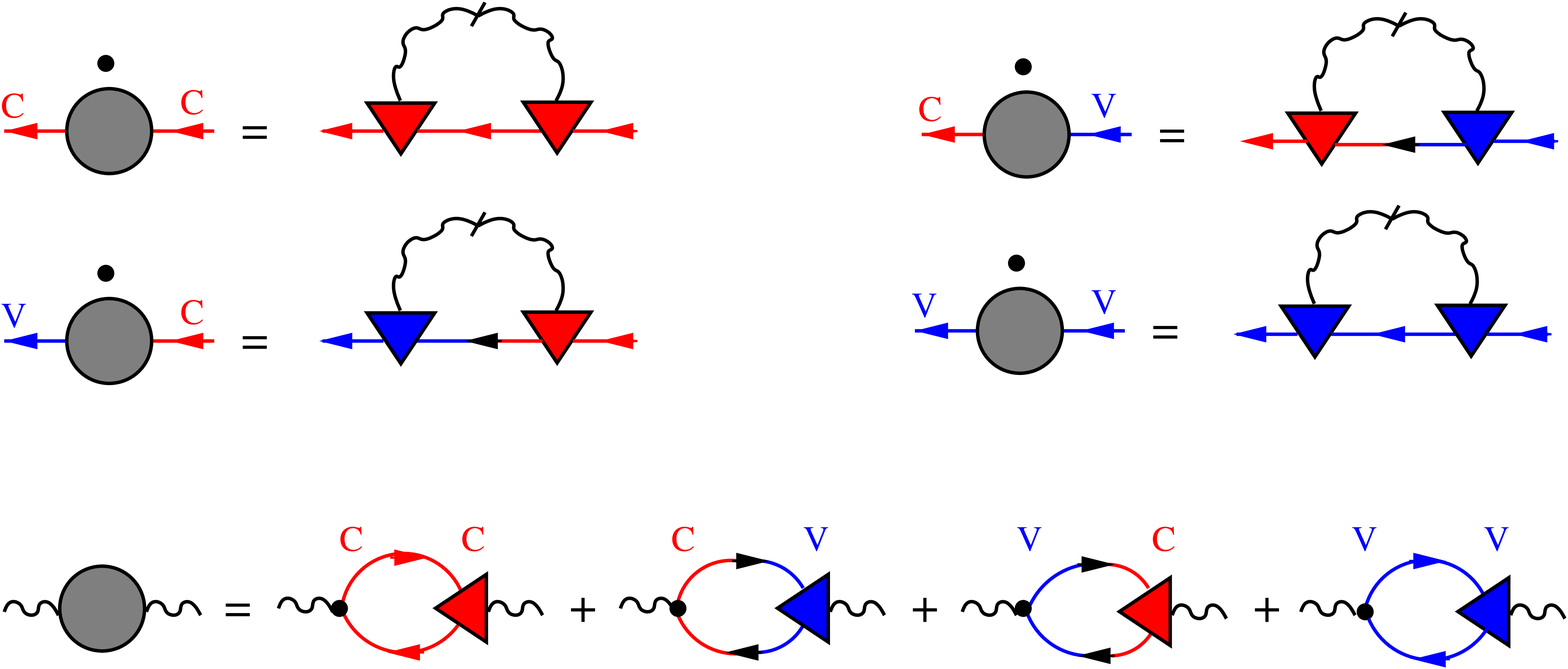}
 \caption{ Upper panel: Graphical illustration of the truncated FRG flow equation for the fermionic self-energy, where the slashed wavy lines are the bosonic single-scale propagators. Lower panel : Exact skeleton equation for the bosonic self-energy. Black dots on the right-hand side denote the bare vertex. All other symbols are the same as in Fig. \ref{fig:fcose}.}
 \label{fig:bcose}
 \end{figure}
A diagrammatic representation of the truncated FRG flow equation for the fermionic self-energy and the exact skeleton equation for the bosonic self-energy is shown in Fig.~\ref{fig:bcose}. 

In order to find the fermionic propagator, we use the fact that our interest lies in the small momentum and low-energy behavior of the model so that we expand the self-energy for small-momentum and energy,
 \begin{equation}
 \Sigma_{\Lambda} (K) = ( 1 - Z_{\Lambda}^{-1} ) i \omega - ( 1 - Y_{\Lambda}^{-1} ) v_F \bd{\sigma} \cdot \bd{k}  + {\cal{O}} ( \omega^2 ),
 \label{eq:sigmalow}
 \end{equation}
and obtain the fermionic propagator using the Dyson equation (\ref{eq:ferprop}),
 \begin{equation}
 G_{\Lambda} (K) = - Z_{\Lambda} \frac{ i \omega + v_{\Lambda} \bd{\sigma} \cdot \bd{k} }{ \omega^2 +  {v}_\Lambda^2 \bd{k}^2 },
 \label{eq:Glow}
 \end{equation}
where the renormalized quasiparticle velocity is given by
 \begin{equation}
 {v}_{\Lambda} = Z_{\Lambda} Y_{\Lambda}^{-1} v_F.
 \label{eq:velbco}
 \end{equation}
 \begin{figure}
 \centering
 \includegraphics[height=5.5cm,width=7.5cm]{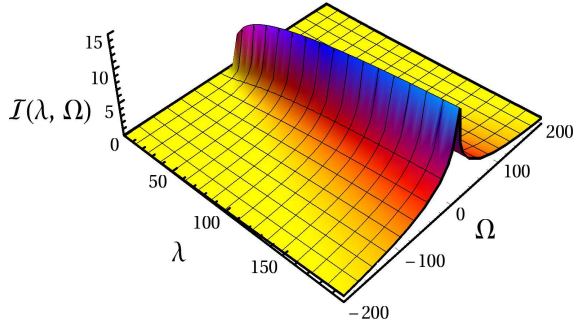}
 \caption{ The integral $\mathcal{I}(\lambda,\Omega)$ as a function of dimensionless momentum $\lambda = \frac{2\Lambda_0}{q}$ and frequency $\Omega = \frac{\bar{\omega}}{v_{\Lambda} q}$, Eq.~(\ref{eq:integral}), showing logarithmic divergence in the limit of small momentum $q$ and frequency $\bar{\omega}$. }
 \label{fig:polarization}
 \end{figure}
On substituting (\ref{eq:Glow}) into (\ref{eq:skeleton2}), performing the frequency integration, and using prolate spheroidal coordinates given in Eq.~(\ref{eq:sphprocor}), we get
\begin{eqnarray}
 \Pi_{\Lambda} (Q) & = &  \frac{\gamma_{\Lambda} Z_{\Lambda}^{2} N_s N_W }{32 \pi^2 } \frac{q^{2}}{v_{\Lambda}} \mathcal{I}(\lambda,\Omega),  
 \label{eq:polarization}
\end{eqnarray}
where
\begin{eqnarray}
 \mathcal{I}(\lambda,\Omega) = \int_{0}^{\pi} d \varphi \sin^3 \varphi \ln\Biggr( \frac{\Omega^2 + (\lambda + \cos \varphi)^2}{1 + \Omega^2} \Biggl),
 \label{eq:integral}
\end{eqnarray}
with $\lambda = \frac{2\Lambda_0}{q}$ and $\Omega = \frac{\bar{\omega}}{v_{\Lambda} q}$ being the dimensionless momentum and frequency. The integration in Eq.~(\ref{eq:integral}) yields a lengthy expression and its plot is shown in Fig.~\ref{fig:polarization} exhibiting a logarithmic divergence in the limit of small momentum $q$ and frequency $\bar{\omega}$. This can be easily seen from the asymptotic limit ($\lambda, \Omega \rightarrow \infty $) of Eq.~(\ref{eq:integral}) which is given by
\begin{eqnarray}
 \mathcal{I}(\lambda,\Omega) \approx  \frac{4}{3} \ln \Biggr( \frac{\lambda^2}{1 + \Omega^2} \Biggl).
 \label{eq:integral2}
\end{eqnarray}
In this limit, the polarization (\ref{eq:polarization}) becomes
\begin{eqnarray}
 \Pi_{\Lambda} (Q) \approx \frac{\gamma_{\Lambda} Z_{\Lambda}^{2} N_s N_W }{24 \pi^2 } \frac{q^{2}}{v_{\Lambda}} \ln \Biggr( \frac{\lambda^2}{1 + \Omega^2} \Biggl).
 \label{eq:polarization2}
\end{eqnarray}

It is important to note that the logarithmically divergent nature of the polarization bubble in 3D is very different compared to 2D graphene where it is linearly proportional to the quasimomentum\cite{Kotov12}. This shows that the Coulomb interaction in 3D Weyl semimetals is strongly screened thereby renormalizing the charge unlike in 2D graphene where the charge does not get renormalized since the polarization bubble goes to zero in the limit of vanishing momentum and energy and the bare Coulomb interaction remains strong and unscreened\cite{Bauer15}.  

Now in order to derive the FRG flow equations for the quasiparticle properties, we return to the self-energy expression, Eq.~(\ref{eq:sigmalow}), from which we obtain the RG flow equations of $Z_{\Lambda}$ and $Y_{\Lambda}$ as
\begin{eqnarray}
 \Lambda \partial_{\Lambda} Z_{\Lambda} & = & \eta_{\Lambda} Z_{\Lambda}, 
 \label{eq:zflowbco}
 \\
\Lambda \partial_{\Lambda} Y_{\Lambda} & = & \tilde{\eta}_{\Lambda} Y_{\Lambda}, 
\label{eq:yflowbco}
 \end{eqnarray}
where the flowing anomalous dimension related to the frequency $\eta_{\Lambda}$ is
 \begin{eqnarray}
\eta_{\Lambda} \ & = & \Lambda Z_{\Lambda} \lim_{ \omega \rightarrow 0} \frac{ \partial}{\partial ( i \omega )} \partial_{\Lambda} \Sigma^{b b}_{\Lambda } ( 0 , i \omega ) \nonumber \\
 & = & Z_{\Lambda}^2 \gamma_{\Lambda}^2 \Lambda \int_Q  \frac{ \delta (q - \Lambda)}{\frac{ \varepsilon_0 \Lambda^{2}}{ 4 \pi e^2} + \Pi_{\Lambda} ( Q ) } \frac{ \bar{\omega}^2 - ( {v}_{\Lambda} q )^2 }{[\bar{\omega}^2 + ( {v}_{\Lambda} q )^2 ]^2 },
 \label{eq:eta2}
 \end{eqnarray}
and the flowing anomalous dimension related to the momentum $\tilde{\eta}_{\Lambda}$ is defined as
 \begin{eqnarray}
 (\bd{\sigma} \cdot {\hat{\bd{k}}} ) \tilde{\eta}_{\Lambda} & = & - \Lambda Y_{\Lambda} \lim_{ | \bd{k} |  \rightarrow 0} \frac{ \partial}{\partial ( v_F | \bd{k} | )} \partial_{\Lambda} \Sigma^{b b^{\prime}}_{\Lambda} (\bd{k} , 0 ),
 \end{eqnarray}
and is given by
 \begin{eqnarray} 
  \tilde{\eta}_{\Lambda} & = & Z_{\Lambda}^2 \gamma_{\Lambda}^2 \Lambda \int_Q \frac{ \delta ( q - \Lambda )}{ \frac{\varepsilon_0 \Lambda^{2}}{ 4 \pi e^2} + \Pi_{\Lambda} ( Q ) } \frac{ \bar{\omega}^2  }{ [  \bar{\omega}^2 + ( {v}_{\Lambda} q )^2 ]^2 }.
 \label{eq:etabar}
 \end{eqnarray}
From Eqs.~(\ref{eq:velbco}), (\ref{eq:zflowbco}), and (\ref{eq:yflowbco}) we get the RG flow of the renormalized velocity,
 \begin{equation}
  \Lambda \partial_{\Lambda} {v}_{\Lambda}  =   ({\eta}_{\Lambda} - \tilde{\eta}_{\Lambda} ) {v}_{\Lambda}. 
  \label{eq:vflowbco}
 \end{equation}
 \begin{figure}
 \centering
 \includegraphics[height=5.5cm,width=8.0cm]{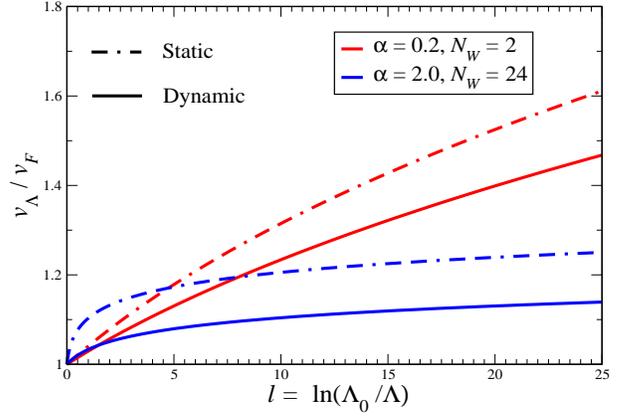}
 \caption{Numerical solution of the dynamical RG flow equation of the velocity, Eq.~(\ref{eq:vflowbco}), for two sets of $\alpha$ and $N_W$. The static results from Eq.~(\ref{eq:flowvel4}), for the same values of $\alpha$ and $N_W$, are also shown as broken lines.}
 \label{fig:vflowbco}
 \end{figure}
 \begin{figure}[b]
 \centering
 \includegraphics[height=5.5cm,width=8.0cm]{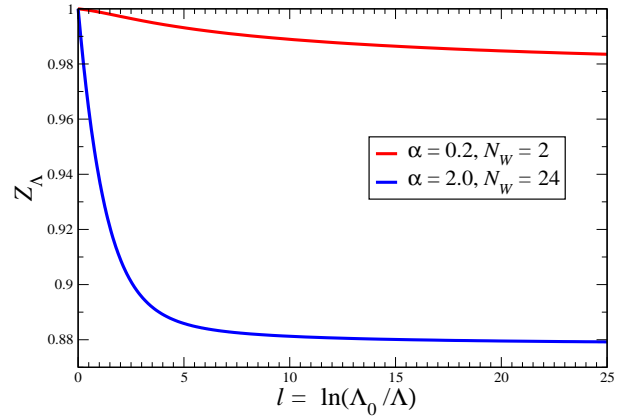}
 \caption{ RG flow of wave function renormalization, $Z_{\Lambda}$, obtained for two sets of $\alpha$ and $N_W$. }
 \label{fig:zflowbco}
 \end{figure}
On using the Ward identity, $ \gamma_{\Lambda} Z_{\Lambda} = 1$, we find that the polarization (\ref{eq:polarization}) gets renormalized only due to renormalized velocity and wave function renormalization. On substituting (\ref{eq:polarization}) into (\ref{eq:eta2}) and (\ref{eq:etabar}) we obtain a closed set of RG flow equations for $ \frac{v_{\Lambda}}{v_F}$ and $Z_{\Lambda}$. On introducing the logarithmic flow parameter $l = \ln \bigl(\frac{\Lambda_0}{\Lambda}\bigr) $ we numerically solve\cite{Note} the coupled integro-differential Eqs.~(\ref{eq:zflowbco}) and (\ref{eq:vflowbco}) using Eqs.~(\ref{eq:eta2}) and (\ref{eq:etabar}). The results for $ \frac{v_{\Lambda}}{v_F}$ and $Z_{\Lambda}$ are shown in Figs.~\ref{fig:vflowbco} and \ref{fig:zflowbco} respectively for two sets of $\alpha$ and $N_W$ which cover the extremes of the parameter range considered in this work. For the sake of comparison, in Fig.~\ref{fig:vflowbco} we also show the static results for the renormalized velocity,  Eq.~(\ref{eq:flowvel4}). 

It is difficult to extract an analytical expression for the renormalized velocity in the dynamic case but on performing a fit we find  that not only the power decreases, compared to Eq.~(\ref{eq:flowvel4}), but also the coefficient in front of the RG parameter reduces. Therefore, in comparison to the static case for given $N_W$ and $\alpha$ the dynamic results for the renormalized velocity get suppressed. This behavior is opposite to the case of 2D graphene where the dynamic interaction enhances the velocity compared to the static approximation. On the other hand, the wave function renormalization in 3D Weyl semimetals flows very slowly towards a finite constant even for large $N_W$ and $\alpha$ and has a finite value at the charge neutral point which suggests that the character of a 3D Weyl fluid is more Fermi-liquid-like with well-defined quasiparticles as in 2D graphene. Our results can have implications on transport\cite{Hosur12} and other physical properties of 3D Weyl semimetals. We also expect our theory to hold for the 3D Dirac semimetals by doubling the Hilbert space and accounting for Kramers degeneracy.

\section{Conclusion and outlook}{\label{sec:conc}}

In summary, we have used a nonperturbative FRG approach to derive and solve the flow equations for the velocity, dielectric function, and the wave function renormalization of clean 3D Weyl semimetals. In order to derive the FRG flow equations for the irreducible vertices of our Hubbard-Stratonovich decoupled low-energy effective model in the presence of static and dynamically screened Coulomb interaction, we have utilized two different cutoff schemes and a truncation based on relevance and symmetry as well as a Ward identity relating the three-legged vertices to the quasiparticle residue. 

In the approximation where the frequency dependence of the dielectric function is neglected, the numerical solutions of our FRG flow equations for the renormalized velocity $v (k)$ and the dielectric function $\varepsilon (q)$ were found to increase with increasing interaction strength $\alpha$ for a fixed number of Weyl nodes $N_W$ while for a given fixed coupling strength the renormalized velocity was found to decrease and the dielectric function increased with increasing number of Weyl nodes $N_W$. We not only obtained the undetermined constants introduced by Abrikosov and Beneslavski{\u{i}}\cite{Abrikosov71}, see Eqs.~(\ref{eq:vel_AB}) and (\ref{eq:de_AB}), given by $a = \frac{2\alpha}{3 \pi} $ and $b = \frac{\alpha N_s N_W}{3 \pi}$ as a function of the number of Weyl nodes and the interaction strength but also found good agreement with their theory as a fit to our numerical solutions.  

We have also included the effects of dynamic screening and evaluated the renormalized velocity and the wave function renormalization. We have shown that for given values of $N_W$ and $\alpha$, the renormalized velocity $v (k)$ is suppressed in the dynamic case as compared to the static approximation. Depending on the values of $N_W$ and $\alpha$, the 3D Weyl semimetals show strong effects of screening due to the charge renormalization, which is absent in graphene, but the wave function renormalization has a finite value at the Weyl node suggesting that the 3D Weyl semimetals behaves like a Fermi-liquid as in 2D graphene. 
 
As an outlook, it would be interesting to study the stability of the phases in the presence of strong coupling and large number of Weyl nodes leading to spontaneous breaking of parity in noncentrosymmetric correlated Weyl semimetals\cite{Sekine14}, excitonic instabilities\cite{Wei14,Gonzalez14,Xue17}, or dynamic mass gap generation\cite{Gonzalez15}, as well as competition between disorder and interactions in 3D topological semimetals\cite{Gonzalez17,Wang17}.

\section*{Acknowledgments}{\label{acknowledgments}}

Two of us (Ar. S. and P. K.) would like to acknowledge the hospitality of the Physics Department at the University of California, Irvine, where part of this work has been carried out.\\

\renewcommand{\appendixname}{APPENDIX}
\appendix*
\section{RG FLOW EQUATIONS OF THE THREE-LEGGED VERTICES}
    
In this appendix, we provide the details of the evaluation of the RG flow equations of the band-label changing mixed fermionic-bosonic, $\Gamma^{(2,1)}$, and the purely bosonic, $\Gamma^{(0,3)}$, three-legged vertices. 

\begin{figure}
 \centering
 \includegraphics[height=5.5cm,width=7.5cm]{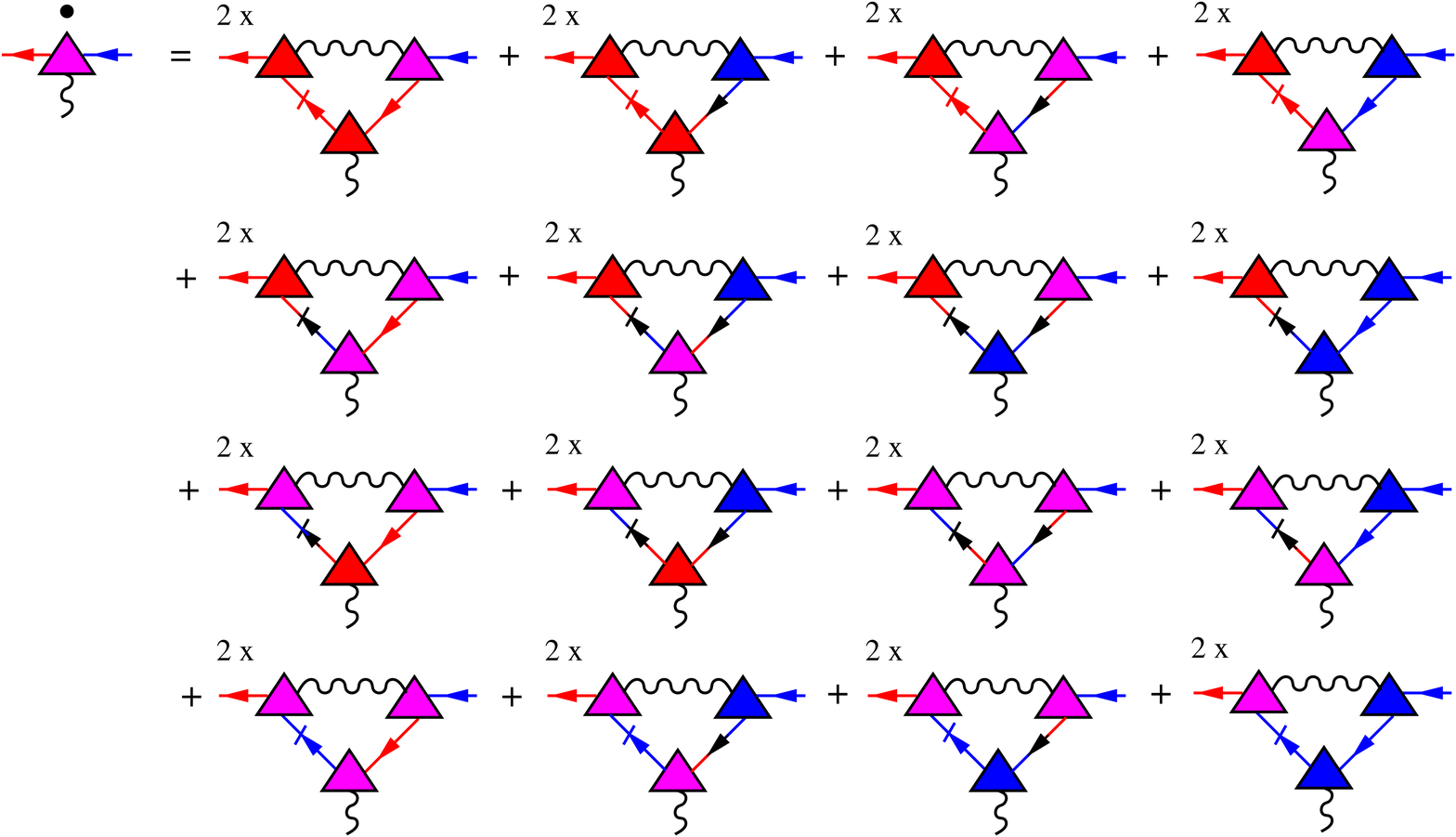}
 \caption{ Diagrammatic representation of the FRG flow equation for the three-legged vertex, Eq.~(\ref{eq:flowvertex3}), with band-label changing two fermionic and one bosonic legs. The magenta triangle signifies the band-label changing three-legged vertex, $\Gamma^{(2,1)}$. The factor 2$\times$ denotes the permutation of the single-scale propagator and the meaning of the other symbols is the same as in Fig. \ref{fig:fcose}.}
 \label{fig:tlfv2}
 \end{figure}

The RG flow of the band-label changing three-legged vertex in the fermionic cutoff scheme is given as
 \begin{widetext}
 \begin{eqnarray} 
 \partial_{\Lambda} \Gamma^{b b^{\prime}}_{n,\Lambda} (K,K-Q,Q) & = & 
 \sum_{b_{1} b_{2} b_{3} b_{4}} \int_{Q^{\prime}} \Bigl\{ F_{\Lambda} ( Q^{\prime}) \Bigl[
 \dot{G}^{b_{1} b_{2}}_{n,\Lambda} (K-Q-Q^{\prime}) G^{b_{3} b_{4}}_{n,\Lambda} (K-Q^{\prime}) + {G}^{b_{1} b_{2}}_{n,\Lambda} (K-Q-Q^{\prime}) \dot{G}^{ b_{3} b_{4} }_{n,\Lambda} (K- Q^{\prime} ) \Bigr]
 \nonumber
 \\
 &  &  \times
  \Gamma^{b b_{1}}_{n,\Lambda} (K,K-Q^{\prime},Q^{\prime}) \Gamma^{b_{2} b_{3}}_{n,\Lambda} (K-Q-Q^{\prime},K-Q,-Q^{\prime}) \Gamma^{b_{4} b^{\prime}}_{n,\Lambda} (K-Q^{\prime},K-Q-Q^{\prime},Q) \Bigr\}, 
 \label{eq:flowvertex3}
 \end{eqnarray}
 \end{widetext}
where $b$ and $b^{\prime}$ are the band indices such that $b \neq b^{\prime}$. The diagrammatic representation of Eq.~(\ref{eq:flowvertex3}) is given in Fig.~\ref{fig:tlfv2} where we show only one of the two band-label changing three-legged vertices. The flow diagram for the other band-label changing three-legged vertex is obtained by interchanging the red and blue lines as well as the corresponding red and blue triangles. Note that in comparison with Eq.~(\ref{eq:flowvertex1}), we have omitted the diagrams consisting of a purely bosonic three-legged vertex in Eq.~(\ref{eq:flowvertex3}) as they will be shown to vanish by symmetry (see below). 
In Eq.~(\ref{eq:flowvertex3}), the momentum and frequency dependence of the three-legged vertices can be neglected so that the integrand becomes a product of the bosonic propagator $F_{\Lambda} (Q)$ and the fermionic propagators, $G_{\Lambda} (Q)$ and $\dot{G}_{\Lambda} (Q)$. Then it becomes sufficient to consider the matrix product $\dot{G}_{\Lambda}(Q)G_{\Lambda}(Q)$ and obtain its off-diagonal components ($b \neq b^{\prime}$). On using Morris lemma (\ref{eq:morrislemma}) and Eqs.~(\ref{eq:glambda}) and (\ref{eq:singlescaleprop}) we get
\begin{equation}
\dot{G}_{\Lambda}(Q)G_{\Lambda}(Q) = \frac{1}{2} \delta(q-\Lambda) \left(\frac{i\bar{\omega}+v_{\Lambda}(q) \bm{\sigma} \cdot \bm{q}}{\bar{\omega}^2 + \xi^2(q)} \right)^2,
\end{equation}
where the off-diagonal components are proportional to the term $\bm{\sigma} \cdot \bm{q}$ and therefore we obtain
\begin{equation}
\left[ \dot{G}_{\Lambda}(Q)G_{\Lambda}(Q)\right]^{b b^{\prime}} = \frac{1}{2} \delta(q-\Lambda) \left[\frac{2i\bar{\omega} v_{\Lambda}(q) \bm{\sigma} \cdot \bm{q}}{\left( \bar{\omega}^2 + \xi^2(q) \right)^2} \right]^{b b^{\prime}},
\label{eq:ggdot}
\end{equation}
which is an odd function of $\bar{\omega}$. But since the bosonic propagator $F_{\Lambda}(Q)=[f_q^{-1}+\Pi_{\Lambda}(Q)]^{-1}$ is an even function of $\bar{\omega}$, as seen from Eq.~(\ref{eq:polflow}), the overall integrand in Eq.~(\ref{eq:flowvertex3}) is an odd function of $\bar{\omega}$ and therefore the frequency integral vanishes. Hence, the band-label changing vertex vanishes for all values of the cutoff. 

In the case of bosonic cutoff, the graphical representation of the band-label changing three-legged vertices is similar to Fig.~\ref{fig:tlfv2} except that instead of the slashed solid line corresponding to the fermionic single-scale propagators, $\dot{G}_{\Lambda}(Q)$, we have a slashed wavy line representing the bosonic single-scale propagator, $\dot{F}_{\Lambda}(Q)$. Thus, the integrand  in Eq.~(\ref{eq:flowvertex3}) is proportional to the product of $\dot{F}_{\Lambda}(Q)$ and the matrix product of two fermionic propagators, $G_{\Lambda} (Q)$. It is straightforward to see that in this case, the off-diagonal components of the matrix product will be proportional to the $\bm{\sigma} \cdot \bm{q}$ term and is an odd function of frequency $\bar{\omega}$ as obtained in Eq.~(\ref{eq:ggdot}). Moreover, since the bosonic self-energy (\ref{eq:polarization}) is an even function of $\bar{\omega}$, the bosonic single-scale propagator (\ref{eq:fdot}) also becomes an even function of $\bar{\omega}$. Therefore, the overall integrand is an odd function of $\bar{\omega}$ and the frequency integral vanishes as in the fermionic cutoff scheme.  

\begin{figure}[!th]
 \centering
 \includegraphics[height=5.5cm,width=6.5cm]{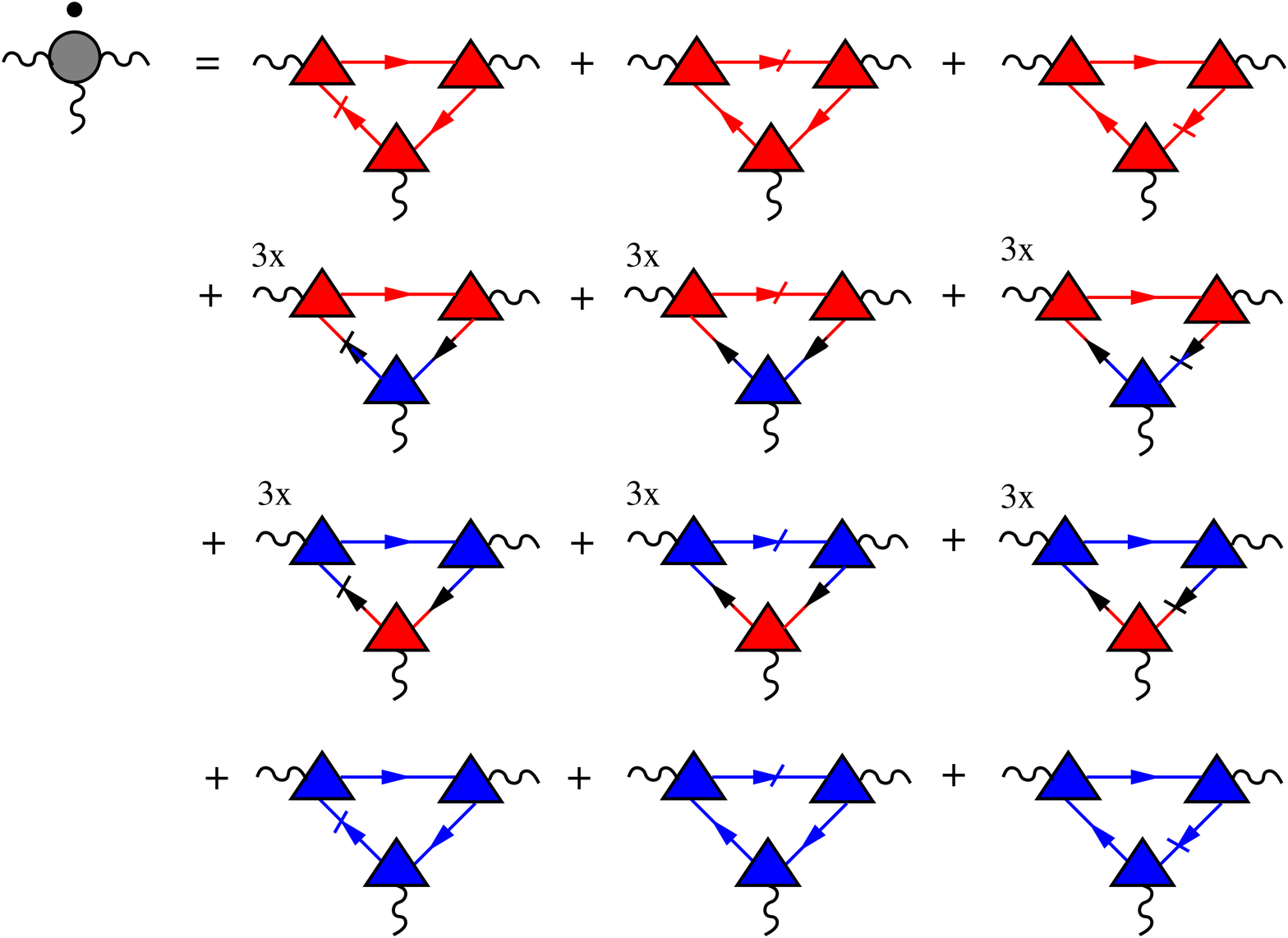}
 \caption{ Graphical representation of the FRG flow equation for the purely bosonic three-legged vertex, Eq.~(\ref{eq:flowbosvertex}). The factor 3$\times$ signifies the cyclic permutation of the vertices in the second and third row for the fixed position of the single-scale propagator. The meaning of the symbols is the same as in Figs. \ref{fig:fcose} and \ref{fig:tlfv}.}
 \label{fig:tlbv}
 \end{figure}

Finally we examine the purely bosonic three-legged vertex. It is generated by the so-called {\textit{triangle diagram}} which is related to the Adler-Bell-Jackiw or chiral gauge anomaly~\cite{Adler69} used to estimate the dc anomalous quantum Hall response. Although such a vertex is absent in the initial action (\ref{eq:Sbare}) and it does not couple to the self-energies within the fermionic cutoff approach, in the following we shall consider its flow equation as given in Eq.~(\ref{eq:flowbosvertex}) and diagrammatically represented in Fig.~\ref{fig:tlbv}. 

Let us neglect the momentum and frequency dependence of the three-legged vertex, $\Gamma^{(0,3)}$, and retain only the marginal part, 
\begin{equation}
\Gamma_{\Lambda}^{(0,3)} (0,0,0) = i \bar{\gamma}_{\Lambda} .
\label{eq:gammabar}
\end{equation}
Therefore, the flow equation of the momentum- and frequency-independent part of the purely three-legged vertex (\ref{eq:flowbosvertex}) becomes
\begin{equation}
\partial_{\Lambda} \bar{\gamma}_{\Lambda} = -3\gamma_{\Lambda}^3 \int_{K} \textrm{Tr} [ \dot{G}_{\Lambda} (K) G_{\Lambda}(K) G_{\Lambda}(K) ],
\label{eq:bosvertflow}
\end{equation}
where we have used Eq.~(\ref{eq:gamma}). On utilizing the Morris lemma (\ref{eq:morrislemma}) along with Eqs.~(\ref{eq:glambda}) and (\ref{eq:singlescaleprop}), we get
\begin{eqnarray}
\partial_{\Lambda} \bar{\gamma}_{\Lambda} & = & \gamma_{\Lambda}^3 \int_{K} \delta (k - \Lambda) \textrm{Tr} \bigl[ [ G^{0} (K) ]^{-1} - \Sigma_{\Lambda} (K) \bigr]^{-3} \nonumber \\
& = & i 2 \gamma_{\Lambda}^3 Z_{\Lambda}^3 \int_{K} \delta (k - \Lambda) \frac{ \omega (\omega^2 - 3\xi_{\Lambda}^2 (k)) }{ \bigl[ \omega^2 + \xi_{\Lambda}^2 (k) \bigr]^{3}} = 0, \nonumber \\
\end{eqnarray}
since the Cauchy principal value of the frequency integral vanishes.

Thus we have shown that the RG flow equations of the band-label changing mixed fermionic-bosonic, $\Gamma^{(2,1)}$, and the purely bosonic, $\Gamma^{(0,3)}$, three-legged vertices vanish identically due to symmetry.


\begin{thebibliography}{99}
\bibitem{Note} The integrals in Eqs.~(\ref{eq:eta2}) and (\ref{eq:etabar}) are convergent over the entire frequency range but due to the logarithmically diverging nature of the polarization bubble, the coupled integro-differential Eqs.~(\ref{eq:zflowbco}) and (\ref{eq:vflowbco}) become non-trivial to solve. We restrict the frequency integral to the limit $\Omega < e^{l}$, as obtained from Eq.~(\ref{eq:polarization2}), and solve the coupled integro-differential equations on a grid.
\end{thebibliography}
\end{document}